\documentclass[twocolumn,article]{autart}  
\usepackage{cite}
\usepackage{bookmark}
\usepackage{amsmath,amssymb,amsfonts}
\usepackage{graphicx}
\usepackage{mathtools}

\usepackage{ntheorem}
\usepackage{epstopdf}
\usepackage{caption}
\usepackage{setspace}
\usepackage{float}
\usepackage{subcaption}
\usepackage{dsfont}
\usepackage{tabularx,ragged2e,booktabs}
\usepackage{algorithm}
\usepackage{algpseudocode}
\makeatletter
\renewcommand*\env@matrix[1][*\c@MaxMatrixCols c]{%
  \hskip -\arraycolsep
  \let\@ifnextchar\new@ifnextchar
  \array{#1}}
\makeatother

\newcommand{\Real}{{\mathds R}} 
\newcommand{\Nat}{{\mathds N}} 

\makeatletter
\newtheorem{definition}{Definition}{}
{}
\newtheorem{proposition}{Proposition}{}
\newtheorem{problem}{Problem}{}
\newtheorem{theorem}{Theorem}{}
\newtheorem{remark}{Remark}{}
{}
{}
\usepackage{pifont}

\usepackage{xcolor}
\hypersetup{
    colorlinks,
    linkcolor={blue!80!black},
    citecolor={blue!80!black},
    urlcolor={blue!10!black}
}
\usepackage{booktabs}

\edef\endfrontmatter{%
  \unexpanded\expandafter{\endfrontmatter}
  \noexpand\endNoHyper 
}
\begin{document}
\begin{frontmatter}
\title{Immersion and Invariance-based Coding \\for Privacy-Preserving Federated Learning\thanksref{footnoteinfo}}
\thanks[footnoteinfo]{This work is supported by European Union’s Horizon Europe programme under grant agreement No 101069748 – SELFY project.}
\author[1]{Haleh Hayati}\ead{h.hayati@tue.nl}
\author[1,2]{Carlos Murguia}\ead{c.g.murguia@tue.nl}
\author[1]{Nathan van de Wouw}\ead{n.v.d.wouw@tue.nl}
\address[1]{Dynamics \& Control Group, Department of Mechanical Engineering, Eindhoven University of Technology, The Netherlands.}
\address[2]{Engineering Systems and Design, Singapore University of Technology and Design, Singapore.}
\begin{abstract} Federated learning (FL) has emerged to preserve privacy in collaborative distributed learning. In FL, clients conduct AI model training directly on their devices rather than sharing their data with a centralized server, which could potentially pose privacy risks. However, it has been shown that despite FL's partial preservation of local data privacy, information about clients' data can still be inferred from shared model updates during the training process. In recent years, several privacy-preserving approaches have been developed to mitigate this privacy leakage in FL. However, they often provide privacy at the cost of model performance or system efficiency. Balancing these trade-offs poses a significant challenge in implementing FL schemes. In this manuscript, we introduce a privacy-preserving FL framework built on the synergy of differential privacy and system immersion and invariance tools from control theory. The core idea is to treat optimization algorithms used in the standard FL schemes (gradient-based algorithms) as a dynamical system that we seek to immerse into a higher-dimensional system (referred here to as the target optimization algorithm). The dynamics of the target optimization algorithm is designed such that, firstly, the model parameters of the original algorithm are immersed/embedded in its parameters, secondly, it works on distorted parameters, and, thirdly, converges to an encoded version of the true model parameters of the original algorithm. The encoded model parameters can be decoded at the server to extract the original model parameters. We demonstrate that the proposed privacy-preserving scheme can be tailored to offer any desired level of differential privacy for local and global model parameters while maintaining the same accuracy and convergence rate as standard FL algorithms.
\end{abstract}
\vspace{-6mm}
\end{frontmatter}
\section{Introduction}\label{sec1}\vspace{-3mm}
\indent Machine learning (ML) has been successfully applied in various applications for multiple fields and industries. 
In traditional machine learning, the server conducting the learning algorithm typically holds the training data centrally. However, when multiple participants are involved, sharing local data with the server poses a significant privacy risk, given the potential disclosure of private information. 
To address this issue, Federated learning (FL) \cite{mcmahan2017communication,li2020federated} has been introduced as a decentralized learning framework, enabling collaboration among numerous participants while preserving data privacy. Its fundamental concept involves training ML models on separate databases distributed across several devices or entities. FL schemes train local models on local clients' databases, with clients subsequently sharing their parameters (e.g., model weights or gradients) with a central server to aggregate a global model. FL is suitable for sensitive data sharing, e.g., in the scope of healthcare and the Internet of Things (IoT), because clients do not need to directly share their training data \cite{shinde2018review}. 
However, despite FL's efforts to preserve clients' raw data privacy by avoiding direct data exchange, research has proven that private information can still be inferred from model parameters throughout the training process. It has been shown that local models can be traced back to their sources \cite{nasr2018comprehensive}. Additionally, private information can be extracted from multiple aggregated global models at the central server \cite{shokri2017membership,so2023securing}. Common attacks to FL are model inversion attacks and gradient inference attacks \cite{fredrikson2015model,aono2017privacy}.\\
\indent In recent years, various approaches have been developed to achieve Privacy-Preserving FL (PPFL) \cite{yin2021comprehensive}. Most of them rely on perturbation-based techniques such as Differential Privacy \cite{abadi2016deep,shokri2015privacy,wei2020federated,liu2024distributed,wang2023decentralized}, and cryptography-based techniques such as Secure Multi-Party Computation \cite{bonawitz2017practical,xu2019verifynet,mohassel2017secureml,mugunthan2019smpai}, and Homomorphic Encryption \cite{ma2022privacy,li2020privacy,asad2020fedopt,lu2018privacy}. Differential Privacy (DP) offers strong probabilistic privacy guarantees with minimal system overhead and algorithmic simplicity. However, it introduces a trade-off between privacy and FL performance, as the added noise can significantly degrade model accuracy and slow convergence. Secure Multi-Party Computation (MPC) allows distributed clients to jointly compute a function without revealing their individual inputs, making it suitable for privacy-preserving model aggregation in FL. While MPC avoids the accuracy loss in DP, it demands additional communication between the server and clients, leading to increased communication costs and overhead. Homomorphic Encryption (HE), another cryptographic method, enables computations on encrypted data. With this approach, clients send their encrypted models to the server to aggregate them without decryption. While HE-based FL preserves model accuracy and eliminates complex client interactions, it is computationally intensive, leading to significant computational overhead. Both cryptographic approaches face challenges related to high communication costs and computational demands. Additionally, while these methods ensure the private aggregation of local models, they are vulnerable to inference attacks over the aggregated models. To address the limitations of cryptographic techniques and DP, recent efforts have focused on hybrid methods combining cryptographic tools with DP schemes to hold acceptable trade-offs between data privacy and FL performance \cite{truex2019hybrid,wang2022decentralized,xu2019hybridalpha,choquette2021capc}.\\
\indent Although existing techniques improve privacy of FL, they often do it at the cost of model performance and system efficiency. Balancing these trade-offs is challenging when implementing private FL systems. Therefore, novel PPFL schemes must be designed to provide strict privacy guarantees with a fair computational cost without compromising the model performance excessively.\\
\indent The aim of this work is to design coding mechanisms that protect the private information of local and global models while implementing FL algorithms. We propose a PPFL framework built on the synergy of random coding and \emph{system immersion} tools \cite{astolfi2003immersion} from control theory. The core idea involves treating the Gradient descent optimization algorithms, e.g., SGD, Adam, Momentum, etc., commonly used in standard FL, as a dynamical system that we seek to \emph{immerse} into a higher-dimensional algorithm (the so-called target optimization algorithm). Essentially, this means that model parameters in the target optimizer must embed all model parameters of the original optimizer (up to a random invertible transformation). The target optimization algorithm must be designed so that: 1) model parameters of the original algorithm are immersed/embedded in its parameters, and 2) it operates on randomly encoded higher-dimensional parameters to produce randomly encoded optimal model parameters. We formulate a coding mechanism at the server side as a random change of coordinates that maps original model parameters to a higher-dimensional parameter space. Such coding enforces that the target optimization algorithm at the clients' side converges to an encoded higher-dimensional version of the optimal model parameters of the standard algorithm. The encoded local model parameters are aggregated by a third party (the so-called aggregator). Another coding is formulated for the aggregator to encode the aggregated model to avoid the server's access to intermediate global models (since the server has access to the first encoding and decoding maps). The aggregator transmits the aggregated encoded global model to the server. The aggregated model is decoded at the server side using the left inverse of the server encoding map (the decoded model at the server side is still encoded by the aggregator coding). Since the aggregator only has access to the encoded local models it does not need to be trusted. In addition, the server only has access to the encoded aggregated model by the aggregator and does not need to be trusted.\\
\begin{table*}[ht]
  \caption{Theoretical comparison of existing privacy-preserving FL frameworks.}
  \centering \scalebox{0.85}{%
  \begin{tabular}{lllll}
    \toprule
    \textbf{Approach} & \textbf{No accuracy loss} &\textbf{No communication costs} & \textbf{No computation overhead}& \textbf{Aggregated model privacy}\\
    \midrule
    DP \cite{abadi2016deep,shokri2015privacy,wei2020federated,liu2024distributed}& \ding{55}&\ding{51}&\ding{51}&\ding{51}\\
    MPC \cite{bonawitz2017practical,xu2019verifynet,mohassel2017secureml,mugunthan2019smpai}&\ding{51}&\ding{55}&\ding{55}&\ding{55} \\
    HE \cite{ma2022privacy,li2020privacy,asad2020fedopt}& \ding{51}&\ding{51}&\ding{55}&\ding{55}\\
    Ours & \ding{51}&\ding{51}&\ding{51}&\ding{51}\\
    \bottomrule
  \end{tabular}}
  \label{tabcompare}
\end{table*}
The general idea of system immersion-based coding is introduced in \cite{hayati2024privacy} as a \emph{homomorphic encryption scheme that operates over the reals} aiming to preserve privacy of centralized 
dynamical algorithms. Compared to standard HE, this approach offers much more freedom to redesign algorithms to work on the encoded/encrypted data. The immersion-based coding scheme provides the same utility as the original algorithm (i.e., when no coding is employed), is computationally efficient, can be applied to large-scale algorithms, and gives arbitrarily strong privacy probabilistic guarantees (in terms of differential privacy) without degrading the algorithm performance. We have started exploring these ideas in \cite{hayati2022privacy,hayati2023mo} to protect privacy of local models in the aggregation steps of FL algorithms. However, it has been shown that the server can recover local models by accessing multiple intermediate global models \cite{so2023securing}. Therefore, it is necessary to provide privacy against inference attacks over both local and intermediate global models. To address this, we propose employing the immersion-based coding idea for privacy of intermediate local and global models in FL by encoding model parameters at both the server and the aggregator at every iteration. This prevents all internal parties involved, server, clients, aggregator, and external parties such as model consumers and eavesdroppers, from accessing actual models. 
We generalize this coding approach for various machine learning and deep learning models to cover gradient descent optimizers for different applications. Simulation experiments and rigorous mathematical proofs indicate that our framework maintains the same accuracy and convergence rate as standard FL, reveals no information about the clients' data, and is computationally efficient.\\
\indent The main contributions of the paper are as follows: 
\vspace{-2mm}
\begin{itemize}
    \item Using random coding and \emph{system immersion} tools, we develop a prescriptive synthesis framework for the design of a privacy-preserving FL algorithm that guarantees data privacy against the inversion of local and global models in the aggregation and broadcasting steps of FL. This extends beyond our previous work in \cite{hayati2022privacy} and other cryptographic tools, which only consider privacy in the aggregation of local models.
    \item We demonstrate that the proposed scheme provides any desired level of differential privacy guarantee for local and global models without compromising the accuracy and convergence rate of the federated learning algorithm with a fair computation cost, in contrast to other related approaches that often impact model performance or system efficiency. This guarantee is not provided in our preliminary work in \cite{hayati2022privacy}.
    \item We validate the effectiveness of the scheme through extensive computer simulations and illustrate that it is possible to employ the proposed scheme to train a variety of ML models without affecting their accuracy and convergence rate through experimental evaluation. Therefore, our analytical results are helpful for the design of privacy-preserving FL architectures with a variety of ML networks and settings.
\end{itemize}
\vspace{-2mm}
Based on the previous discussion, summarized in Table \ref{tabcompare}, our proposed PPFL approach is advantageous in terms of model performance, system efficiency, and privacy of intermediate aggregated models.\\
The notation of the paper is introduced in Table \ref{tabnotations}. 
\vspace{-2mm}
\begin{table}[t]
  \caption{Description of Main Notation.}
  \centering \scalebox{0.9}{%
  \begin{tabular}{ll}
    \toprule
    \textbf{Symbol} & \textbf{Description} \\
    \midrule
    $K$ & Number of local iterations for clients\\
    $T$ & Number of global iterations of FL\\
    $N_c$& Number of clients\\
    $\mathcal{D}$, $\mathcal{D}^{\prime}$ & Adjacent databases\\
    $\mathcal{C}_i$ &  $i^{th}$ client\\
    $\mathcal{D}_i$ & The database held by $\mathcal{C}_i$ \\
    $w_{i,{k}}$ & $\mathcal{C}_i$ 's local parameters at $k^{t h}$ local iteration\\
    $w_{i,{K}}$ & Local uploading parameters of client $\mathcal{C}_i$\\
    $w^t$& Global aggregated parameters from all local \\
    ~ &parameters
at the $t^{th}$ global iteration\\
$\tilde{w}^t$& Global aggregated parameters encoded by\\
    ~ & the server at $t^{th}$ global iteration\\
$\bar{w}^t$& Global aggregated parameters encoded by \\ ~ &the aggregator at $t^{th}$ global iteration\\
${{w}^{\prime}}^t$& Global aggregated parameters encoded by the\\ ~ &aggregator and server at $t^{th}$ global iteration\\
$\pi_1(\cdot)$& Server encoding map\\
$\pi^L_1(\cdot)$& Server decoding map (left inverse of $\pi_1(\cdot)$)\\
$\pi_2(\cdot)$& Aggregator encoding map\\
$\pi^R_2(\cdot)$& Clients decoding map (right inverse of $\pi_2(\cdot)$)\\
$n$& Number of model parameters\\
$\tilde{n}$& Number of parameters encoded by server\\
$\bar{n}$ & Number of parameters encoded by aggregator\\
${n}^{\prime}$ & Number of parameters after encoding\\ ~ &by aggregator and server\\
    \bottomrule
  \end{tabular}}
  \label{tabnotations}
\end{table}
\vspace{-2mm}
\section{Problem Formulation}\label{sec2}
\vspace{-2mm}
\subsection{Standard Federated Learning}
\vspace{-2mm}
Our scheme's architecture is developed by expanding upon the standard FL algorithm. Within the standard FL framework, the objective is to train a global AI model collaboratively across multiple dispersed devices, referred to as clients, and a central server while avoiding the direct exchange of local data held by the clients. Instead, clients transmit their local model parameters to the server, obtained through training on their devices with local data. The server then aggregates these local models' parameters to construct a comprehensive global model, which is subsequently sent back to the clients. These updated global parameters serve as the initial conditions for clients to refine their local models. This iterative process continues until convergence \cite{konevcny2016federated}.\\
\indent We consider a standard FL system consisting of one server and $N_c$ clients. The local database owned by the $i^th$ client, $\mathcal{C}_i$, $i \in\{1,2, \ldots, N_c\}$, is denoted by $\mathcal{D}_{i}$. At each iteration $t \in \mathbb{N}$, the server broadcasts the latest global model, $w^t \in \mathbb{R}^{n}$ (a vector of parameters), to all clients (beginning from a random initial model $w^0$). These iteration times $t$ are termed as global iterations. Next, clients use the latest update on $w^t$ and local data $\mathcal{D}_{i}$ to minimize a specified loss function $l(w_i^t,\mathcal{D}_i)$ at their devices to identify local AI models, $w_i^t \in \mathbb{R}^{n}$. This can be written as follows:
\begin{equation}
    w_i^{t+1}=\arg \min _{{w}_{i}^{t}} l\left({w}_{i}^{t},\mathcal{D}_i\right). \label{lossfunc}
\end{equation}
The clients transmit their local optimal $w_i^{t+1}$ back to the server, which then proceeds to update the global model as follows:
\begin{equation}
  {w}^{t+1}=\sum_{i=1}^{N_c} \frac{\left|\mathcal{D}_{i}\right|}{|\mathcal{D}|} {w}^{t+1}_{i},\label{serveraggregate}
\end{equation}
where $|\mathcal{D}_i|$ is the size of the $i^{th}$ database, $|\mathcal{D}| := \sum_i |\mathcal{D}_i|$, and ${w^t}$ is the global aggregated model. The procedure iterates until reaching convergence towards the global optimum (see \cite{mcmahan2017communication}):
\begin{equation}\label{global}
    {w}^{*}=\arg \min _{w} \sum_{i=1}^{N_c}\frac{\left|\mathcal{D}_{i}\right|}{|\mathcal{D}|} l\left(w,\mathcal{D}_{i}\right).
\end{equation}
\indent In general, standard FL clients use gradient descent optimization algorithms \cite{ruder2016overview}, e.g., Stochastic Gradient Descent (SGD), Adam, etc., as the optimization tool to minimize their local loss function \eqref{lossfunc}. Each client calculates the stochastic gradient of the local model using their local data $\mathcal{D}_{i}$ and updates its local model following $K$ iterations of the optimizers that can generally be modeled as follows:
\begin{equation}
\text{Optimizer} \left\{
\begin{aligned}
&w_{i,0}=w^{t}, \\
&w_{i,{k+1}} = f({w}_{i,k},\mathcal{D}_{i}):={w}_{i,k} -
g({w}_{i,k},\mathcal{D}_{i}),  \label{standardoptimizer}\\
&k=0,1, \cdots, K-1,\\
&w_{i}^{t+1} = w_{i,K},
\end{aligned}\right.
\end{equation}
where $w_{i,{k}} \in \mathbb{R}^{n}$ denotes the local model parameters in the $k^{th}$ local iteration of the optimization algorithm at the $i^{th}$ client, and $g({w}_{i,k},\mathcal{D}_{i})$ is a gradient-based function, reflecting the step in parameters, that can be defined for each gradient descent optimizer.\\
Stochastic Gradient Descent (SGD) is a common gradient descent optimization technique in FL \cite{ruder2016overview}. For SGD, function $g(\cdot,\cdot)$ can be written as $g({w}_{i,k},\mathcal{D}_{i}):=\eta \nabla {l}({w}_{i,k},\mathcal{X}_{i,k})$, where $\eta>0$ is the learning rate, $\nabla {l}(\cdot,\cdot)$ is the gradient of the loss function ${l}(\cdot,\cdot)$, and $\mathcal{X}_{i,k}$ is a minibatch of the database (a subset of the local database $\mathcal{D}_{i}$) in the $k^{th}$ iteration.\\
\indent At every round, each client initializes a gradient descent optimization algorithm using the latest received $w^t$ and updates $w_{i}^{t+1}$ via $K$ iterations of the optimizer \eqref{standardoptimizer}, i.e., $w_{i}^{t+1} = w_{i,K}$. Optimal local parameters, $w_{i}^{t+1}$, are sent to the server for aggregation. After a sufficient number of global iterations between clients and the server (in the global counter $t$) and local updates (in the local counter $k$), the standard FL scheme converges to the optimal global model \eqref{global} (see \cite{mcmahan2017communication}).
\vspace{-1mm}
\subsection{Privacy Requirements}
\vspace{-2mm}
Information about clients' private data can be inferred from the model updates throughout the training process \cite{shokri2017membership,nasr2018comprehensive,fredrikson2015model,aono2017privacy}. In addition, information leakages can also occur in the broadcasting step by analyzing the global model parameters \cite{shokri2017membership}. In FL, two types of actors can infer private information: internal actors (participating clients, the central server, and third parties) and external actors (model consumers and eavesdroppers) \cite{yin2021comprehensive}. We assume all the internal actors are untrusted (honest-but-curious), which means they will faithfully follow the designed FL protocol but attempt to infer private information. External actors are also untrusted; they aim to eavesdrop the communication between internal actors to infer information.\\
\indent To address the problem of deducing private information of clients from their uploaded local models, numerous cryptographic privacy-preserving methods such as MPC and HE and perturbation-based techniques such as DP are usually employed to ensure that clients’ local models are not accessed by the server or any malicious actors. Generally, these methods are designed to ensure that local models are not directly exposed to other parties to prevent inferring sensitive data.\\ 
\indent Unfortunately, approaches exclusively using cryptographic methods remain vulnerable to inference over the aggregated models. As the aggregated models in every iteration remain unchanged from function execution without privacy, it has been shown that the server can recover local models through multiple intermediate global models \cite{so2023securing}. Therefore, we must also consider potential inference over the aggregated models. Solutions addressing the aggregated models' privacy mostly use a DP framework. However, there is a trade-off between DP and the performance of FL, both in terms of model accuracy and convergence rate due to the added noises.\\
\indent In this study, we concentrate on protecting privacy against inference over intermediate local and global models, in the aggregation and broadcasting steps of FL without degrading the accuracy and convergence rate. 
\vspace{-1mm}
\subsection{Privacy-preserving FL Problem}
\vspace{-2mm}
\indent To prevent inference of the clients' databases from their local updates, we propose a privacy-preserving FL scheme to distort local updates $w_i^t$ before transmission. Starting from the initial local model of clients $w_{i,0}$, which is the latest broadcasted global model by the server, $w_{i,0}=w^t$, we let the server distort the original aggregated update before disclosure through some encoding map $\pi_1:\mathbb{R}^{n} \rightarrow \mathbb{R}^{{\tilde{n}}}$, $\tilde{w}^t=\pi_1({w}^t)$, and send the distorted $\tilde{w}^t$ to the clients to run the optimization algorithm and update their local models. In general, running the standard optimization algorithm \eqref{standardoptimizer} on the distorted initial model $\tilde{w}_{i,0}=\tilde{w}^t$ will not yield the same model update ${w}_{i,K}$ that would be obtained if it was run using ${w}^t$. That is privacy-preserving FL methods that do not account for (remove) the distortion induced by the encoding map $\pi_1(\cdot)$ lead to performance degradation. This is precisely the problem with perturbation-based techniques for privacy preservation (like with standard, non-homomorphic DP tools).\\
\indent To address these challenges, the scheme proposed here seeks to design a new gradient descent optimization algorithm (referred hereafter to as the target optimizer) that runs on encoded model from the server $\tilde{w}^t$ and returns an encoded local model update, $\Tilde{w}_i^{t+1}$, that can be later decoded after aggregating local updates at the server side. Note that, we are not looking for one particular target algorithm but a methodology that can construct such target algorithm for any (or a broad class of) original optimization algorithm.\\
\indent We seek to design $\pi_1(\cdot)$ such that $\tilde{w}^t=\pi_1({w}^t) \in \mathbb{R}^{\tilde{n}}$ is of higher dimension than $w^t \in \mathbb{R}^{n}$, i.e., $\tilde{n} > n$. We impose this condition to create redundancy in both the encoding map and target optimizer. Redundancy will allow us to inject randomness that can be traced through the algorithm, removed after model aggregation, and used to enforce an arbitrary level of differential privacy. Consider the higher-dimensional target optimizer:
\begin{equation}
\text{Target optimizer} \left\{
\begin{aligned}
&\tilde{w}_{i,0}=\tilde{w}^{t}, \\
&\tilde{w}_{i,{k+1}}= \tilde{f}\left(\tilde{w}_{i,k},\mathcal{D}_{i}\right),\label{targetoptimizerstep}\\
&k=0,1, \cdots, K-1,\\
&\tilde{w}_{i}^{t+1} = \tilde{w}_{i,K},
\end{aligned}\right.
\end{equation}
with function $\tilde{f}:\mathbb{R}^{\Tilde{n}} \rightarrow \mathbb{R}^{\Tilde{n}}$, $\tilde{n}_y > n_y$, to be designed, distorted initial model parameters $\tilde{w}^{t}$ (the latest encoded global update from the server), and distorted local update $\tilde{w}_{i}^{t+1}$ generated by the target optimizer.\\
\indent Our goal is to design the encoding map $\pi_1(\cdot)$ and the functions $\Tilde{f}(\cdot)$ such that the target optimizer can work on the encoded data $\tilde{w}_{i,0}=\tilde{w}^t$ to produce encoded model update $\tilde{w}_i^{t+1}$ that can be used to extract ${w}^{t+1}$ after aggregation. Note that the choice of $\tilde{f}(\cdot)$ in \eqref{targetoptimizerstep} provides a prescriptive design in terms of the standard optimization function $f(\cdot)$ in \eqref{standardoptimizer}.\\ 
\indent In our setting, we consider a third party other than the clients and the server for model aggregation. We refer to this party simply as the \emph{aggregator}. We consider that once a complete cycle has been finished at every client by the target optimizer, all clients send their last iteration, $\tilde{w}^{t+1}_i=\tilde{w}_{i,K}$, to the aggregator for model aggregation. The role of the aggregator is to interface between clients and the server and thus prevent the server from accessing exact encoded local models since it has access to the encoding map $\pi_1(\cdot)$. The aggregator takes the updated encoded local models from all clients, $\tilde{w}^{t+1}_i$, aggregates them, and sends (only) the aggregated encoded model to the server. Hence, the server cannot access any local model and only has access to the aggregated results. Since the aggregator only accesses the encoded local updates $\tilde{w}^{t+1}_{i}$ and does not have access to $\pi_1(\cdot)$, it is not required to be trusted.\\ 
\indent At a system-theoretic level, what we seek to accomplish is to embed model parameters ${w}_{i,k}$ of the standard optimizer (initialized with the latest model from the server ${w}^t$) into model parameters $\tilde{w}_{i,k}$ of the target optimizer (initialized with the latest encoded model from the server $\tilde{w}^t$)). That is, we aim to design $\pi_1(\cdot)$ and the target optimizer so that there exists a bijection between model parameters of both optimizers (referred here to as the \emph{immersion map}), and thus, having model parameters of the target optimizer uniquely determining the model parameters of the standard one through the immersion map. This leads to the possibility of running the target optimizer (instead of the standard optimizer) by clients on encoded parameters $\tilde{w}^t$, and then, aggregating their local updates $\tilde{w}_i^{t+1}$ by the aggregator to achieve aggregated encoded model $\tilde{w}^{t+1}$, which can be used to extract the exact aggregated global model ${w}^{t+1}$ from $\tilde{w}^{t+1}$ by the server. Hence, a fundamental question is how do we design $\pi_1(\cdot)$ and the target optimizer $\tilde{f}(\cdot)$ in \eqref{targetoptimizerstep} to accomplish this bijection?\\
\indent In system and control theory, this type of embedding between systems trajectories is referred to as \emph{system immersion} and has been used for nonlinear adaptive control \cite{astolfi2003immersion,AstolfiBook} and output regulation \cite{IsidoriRegulation,IsidoriBook}. Using system immersion tools in the context of privacy in cloud computing has been studied in \cite{hayati2024privacy}. In what follows, we explore the idea of using system immersion tools to design a privacy-preserving federated learning algorithm. We develop the necessary mathematical machinery and provide sufficient conditions to simultaneously design the encoding map $\pi_1(\cdot)$ and the target optimizer $\tilde{f}(\cdot)$ to accomplish immersion and aggregated model extraction using ideas from system immersion. This will culminate in a problem description on immersion-based coding for privacy-preserving FL at the end of this section.
\subsection{Immersion-based Privacy-preserving FL: Secure Aggregation}
Consider the standard and target optimization algorithms in \eqref{standardoptimizer} and \eqref{targetoptimizerstep}, respectively. We say that standard optimizer is immersed in target optimizer if $\tilde{n} > {n}$ and there exists a function $\pi_1: \mathbb{R}^{n} \rightarrow\mathbb{R}^{\tilde{n}}$ that satisfies $\tilde{w}_{i,k}=\pi_1({w}_{i,k})$ for all ${w}_{i,k}$ and $\tilde{w}_{i,k}$ generated by the optimizer and target optimizer, respectively. That is, any model parameters of the target optimizer are model parameters of the original optimizer through the mapping $\pi_1(\cdot)$, and $\pi_1(\cdot)$ is an immersion because the dimension of its image is ${\tilde{n}}>{{n}}$.
We refer to this map $\pi_1(\cdot)$ as the \emph{immersion map}.\\
\indent To guarantee that the standard optimizer \eqref{standardoptimizer} is immersed in the target optimizer \eqref{targetoptimizerstep} (in the sense introduced above), we need to impose conditions on the functions shaping the optimizers, their initial conditions, and the immersion map, i.e., on $(f,{w}_{i,0},\pi_1, \tilde{f},\tilde{w}_{i,0})$. In particular, we require to design $(\pi_1, \tilde{f},\tilde{w}_{i,0})$ such that $\tilde{w}_{i,k}=\pi_1({w}_{i,k})$ for all $k$, i.e., the manifold $\tilde{w}_{i,k}=\pi_1({w}_{i,k})$ must be forward invariant under the optimization algorithms in \eqref{standardoptimizer} and \eqref{targetoptimizerstep} \cite{haro2016parameterization}. Let us define the off-the-manifold error $e_k:=\tilde{w}_{i,k}-\pi_1({w}_{i,k})$. The manifold $\tilde{w}_{i,k}=\pi_1({w}_{i,k})$ is forward invariant if and only if the origin of the error dynamics:
\begin{equation}
\begin{aligned}\label{errordynamic}
&{e}_{k+1}=\tilde{w}_{i,{k+1}}-\pi_1({w}_{i,{k+1}})\\
&=\tilde{f}(e_k+\pi_1({w}_{i,{k}}),\mathcal{D}_{i})-\pi_1(f({w}_{i,{k}},\mathcal{D}_{i})),
\end{aligned}
\end{equation}
is a fixed point, i.e., $e_k=\mathbf{0}$ implies $e_{k+1}=\mathbf{0}$ for all $k \in  \Nat_0$ \cite{hirsch2012differential}. Substituting $e_k=\mathbf{0}$ and $e_{k+1}=\mathbf{0}$ in \eqref{errordynamic} leads to
\begin{equation} \label{dynamicsmap}
    \tilde{f}(\pi_1({w}_{i,{k}}),\mathcal{D}_{i})-\pi_1(f(({w}_{i,{k}}),\mathcal{D}_{i}))=\mathbf{0}.
\end{equation}
Therefore, $\tilde{w}_{i,k}=\pi_1({w}_{i,k})$ is satisfied for all $k$ if: \textbf{(1)} the initial condition of \eqref{targetoptimizerstep}, $\tilde{w}_{i,0}$, satisfies $\tilde{w}_{i,0} = \pi_1({w}_{i,0})$, which leads to $e_0=\mathbf{0}$ (start on the manifold); and \textbf{(2)} the dynamics of both algorithms match under the immersion map, i.e., \eqref{dynamicsmap} is satisfied (invariance condition on the manifold)  for all $k \geq 0$. We refer to these two conditions as the \emph{immersion conditions}.
\\[1mm]
\noindent\makebox[\linewidth]{\rule{\linewidth}{0.8pt}}
\textbf{Immersion Conditions:}\
\begin{equation}\label{immersionconditiongeneral}
    \left\{
\begin{aligned}
&\tilde{f}(\pi_1({w}_{i,{k}}),\mathcal{D}_{i})=\pi_1(f(({w}_{i,{k}}),\mathcal{D}_{i})), \text{ (invariance)}\\
&\tilde{w}_{i,0} = \pi_1({w}_{i,0}). \text { (start on the manifold)}
\end{aligned}
\right. \vspace{-2.5mm}
\end{equation}
\noindent\makebox[\linewidth]{\rule{\linewidth}{0.8pt}}
The start on the manifold condition is satisfied by encoding the aggregated global model by the server through the mapping $\pi_1(\cdot)$ at every global iteration as $\Tilde{w}^t=\pi_1({w}^t)$. The server broadcasts the encoded aggregated model $\Tilde{w}^t$ to all clients. Then, each client initializes their target optimizer using the latest received encoded global model as $\Tilde{w}_{i,0}=\pi_1({w}_{i,0})=\Tilde{w}^t$ and updates $\Tilde{w}_{i}^{t+1}$ via $K$ iterations of the target optimizer \eqref{targetoptimizerstep}, i.e., $\Tilde{w}_{i}^{t+1} = \Tilde{w}_{i,K}$.\\
\indent So far, we have derived sufficient conditions \eqref{immersionconditiongeneral} for local model parameters of the standard optimizer to be immersed into the model parameters of the target optimizer in terms of $(f,\tilde{f},{w}_{i,0},\Tilde{w}_{i,0},\pi_1)$. Next, we derive conditions on the immersion map $\pi_1(\cdot)$ so that the encoded aggregated model by the aggregator, $\Tilde{w}_a^{t+1}$, can be decoded to extract the original aggregated model ${w}^{t+1}$ by the server. The aggregated model by the aggregator at iteration $t$ is given by
\begin{equation}
   \tilde{w}_a^{t+1}=\sum_{i=1}^{N_c} \frac{|\mathcal{D}_i|}{|\mathcal{D}|} \tilde{w}^{t+1}_{i} = \sum_{i=1}^{N_c} \frac{|\mathcal{D}_i|}{|\mathcal{D}|} \pi_1\left(w^{t+1}_{i}\right) \label{distortedaggregatedModel},
\end{equation}
where the right-hand side part of \eqref{distortedaggregatedModel} follows from the immersion condition $\tilde{w}_{i,K}=\pi_1({w}_{i,K})$.\\
\indent The server receives $\tilde{w}^{t+1}_a$ in \eqref{distortedaggregatedModel} and aims to retrieve $w^{t+1} = \sum_{i=1}^{N_c} (|\mathcal{D}_i|/|\mathcal{D}|) w^k_{i}$ -- the aggregated result of the standard optimization algorithm in \eqref{standardoptimizer}. The latter imposes an extra condition on the immersion map, $\pi_1(\cdot)$, since to retrieve $w^{t+1}$ from $\tilde{w}^{t+1}$, there must exist a left-inverse function $\pi_1^L:\mathbb{R}^{\Tilde{n}} \to \mathbb{R}^n$ of $\pi_1(\cdot)$, i.e., satisfying the following left-invertibility condition $\pi_1^L \left( \tilde{w}^{t+1}\right) = {w}^{t+1}$:
\begin{equation}\label{left_inverse}
\pi_1^L \left( \tilde{w}^{t+1}\right)=\pi_1^L \left( \sum_{i=1}^{N_c} \frac{|\mathcal{D}_i|}{|\mathcal{D}|} \pi_1\left(w^{t+1}_{i}\right) \right) = \sum_{i=1}^{N_c} \frac{|\mathcal{D}_i|}{|\mathcal{D}|} w^{t+1}_{i}.
\end{equation}
If such $\pi_1^L(\cdot)$ and $\pi_1(\cdot)$ exist, the server can retrieve the original aggregated parameters $w^{t+1}=\sum_{i=1}^{N_c} (|\mathcal{D}_i|/|\mathcal{D}|) w^{t+1}_i$ by passing the encoded results through the function $\pi^L(\cdot)$. We now have all the machinery to state the problem we seek to solve.

\begin{problem}\label{problem1}\emph{\textbf{(Immersion-based Privacy-Preserving FL)}} For given $(f,w_{i,0})$ of the standard optimizer \eqref{standardoptimizer}, design an immersion map $\pi_1(\cdot)$, and $(\tilde{f},\tilde{w}_{i,0})$ of the target optimizer \eqref{targetoptimizerstep} so that: \textbf{(a)} the immersion conditions \eqref{immersionconditiongeneral} hold; and \textbf{(b)}  there exists a function $\pi^L(\cdot)$ satisfying \eqref{left_inverse}.
\end{problem}
\begin{remark}
Solutions to Problem \ref{problem1} characterize a class of encoding maps and target optimizers for which we can design homomorphic encryption schemes (a prescriptive design for given $(f,w_{i,0})$). However, this class is infinite-dimensional (over a function space). It leads to an underdetermined algebraic problem with an infinite-dimensional solution space. To address this aspect, we impose structure on the maps we seek to design. We restrict to random affine maps composed of linear coordinate transformations and additive random processes. In what follows, we prove that this class of maps is sufficient to guarantee an arbitrary level of differential privacy.
\end{remark}
\section{Immersion-based Coding for Privacy-
Preserving FL}\label{sec3}
\vspace{-1mm}
In this section, we construct a prescriptive solution to Problem \ref{problem1} using a random affine immersion map $\pi_1(\cdot)$. As the problem formulation and solution are based on the system immersion theory, we refer to our algorithm as \emph{System Immersion based Federated Learning} (SIFL).\\
\indent Let the immersion map $\pi_1(\cdot)$ be an affine function of the form:
\begin{equation}\label{privacymechanism2}
\pi_1\left(w_{i,k}\right) := \Pi_1 w_{i,k} +b_1^t,
\end{equation}
for matrix $\Pi_1 \in \mathbb{R}^{\tilde{n} \times n}$, $\tilde{n}> n$, to be designed, and some i.i.d. multivariate random process $b_1^t \in \mathbb{R}^{\tilde{n}}$. For this map, the immersion condition \eqref{immersionconditiongeneral} amounts to $\tilde{w}_{i,0} = \pi_1({w}_{i,0})$ (encoded by the server before broadcasting) and
\begin{equation}\label{immersioncondition2}
\tilde{f}(\Pi_1{w}_{i,{k}}+b_1^t,\mathcal{D}_{i})=\Pi_1 f(({w}_{i,{k}}),\mathcal{D}_{i})+b_1^t.
\end{equation}
Let the function $\tilde{f}(\cdot)$ be designed in the following form (using the form of the original optimizer in \eqref{standardoptimizer}):
\begin{equation}\label{eq:20}
\tilde{f}(\tilde{w}_{i,k},\mathcal{D}_{i}):=\tilde{w}_{i,k} -
M_2 g( M_1 \tilde{w}_{i,k},\mathcal{D}_{i}),
\end{equation}
for some matrices $M_1 \in \mathbb{R}^{n \times \tilde{n}}$ and $M_2 \in \mathbb{R}^{\tilde{n} \times n}$ to be designed. Hence, the immersion condition \eqref{immersioncondition2} takes the form:
\begin{equation}\label{eqc1}
\begin{aligned}
\Pi_1 {w}_{i,{k}}+b_1^t -
&M_2 g( M_1(\Pi_1 {w}_{i,{k}} +b_1^t),\mathcal{D}_{i})\\& \quad = \Pi_1({w}_{i,{k}} -
 g( {w}_{i,{k}},\mathcal{D}_{i})) + b_1^t.
 \end{aligned}
\end{equation}
Note that the choice of $\tilde{f}(\cdot)$ in \eqref{eq:20} provides a prescriptive design in terms of the standard optimization function $f(\cdot)$ in \eqref{standardoptimizer}. That is, we exploit the knowledge of the original optimizer and build the target optimizer on top of it in an algebraic manner.\\
\indent To satisfy \eqref{eqc1}, we must enforce $M_1(\Pi_1 {w}_{i,{k}} +b_1^t)={w}_{i,{k}}$ and $M_2 =\Pi_1$, which implies that $M_1 \Pi_1=I$, $M_1 b_1^t=\mathbf{0}$ (i.e., $b_1^t \in \text{ker}[M_1]$). From this brief analysis, we can draw the following conclusions: 1) $\Pi_1$ \emph{must be of full column rank} (i.e., $\text{rank}[\Pi_1]=n$); 2) $M_1$ is the left inverse of $\Pi_1$, i.e., $M_1 = \Pi^L_1$ (which always exists given the rank of $\Pi_1$); 3) $b_1^t \in \text{ker}[\Pi^L_1]$ (this kernel is always nontrivial because $\Pi_1$ is full column rank by construction); and 4) $M_2 =\Pi_1$. Combining all these facts, the final form of $\tilde{f}(\cdot)$ in \eqref{targetoptimizerstep} is given by
\begin{equation}\label{eqc2}
\tilde{f}(\tilde{w}_{i,k},\mathcal{D}_{i}):=\tilde{w}_{i,k} -
\Pi_1 g( \Pi^L_1 \tilde{w}_{i,k},\mathcal{D}_{i}).\end{equation}
At every global iteration $t$, vector $b_1^t$ is designed by the server to satisfy $\Pi_{1}^{L} b_1^t=\mathbf{0}$ and used to distort the coding map $\pi_1(\cdot)$ in \eqref{privacymechanism2}. The role of $b_1^t$ is to randomize the model parameters so that we can guarantee differential privacy. The idea is that because the server (that applies mapping $\pi_1(\cdot)$ to the global model before sending it to clients) knows that $b_1^t$ draws realizations from the kernel of $\Pi_{1}^{L}$, the distortion induced by it can be removed at the server side after aggregation by the aggregator. 
To enforce that $b_1^t \in \text{ker}[\Pi_{1}^{L}]$, without loss of generality, we let it be of the form $b_1^t=N_1 r_1^t$ for some matrix $N_1 \in \mathbb{R}^{\tilde{n} \times (\tilde{n} - n)}$ expanding the kernel of $\Pi_{1}^{L}$ (i.e., $\Pi_{1}^{L}N_1=\mathbf{0}$) and an arbitrary i.i.d. process $r_1^t \in \mathbb{R}^{(\tilde{n} - n)}$. This structure for $b_1^t$ always satisfies $\Pi_{1}^{L} b_1^t = \Pi_{1}^{L}N_1  r_1^t = \mathbf{0}$, for $ r_1^t$ with arbitrary probability distribution.\\
\indent So far, we have designed $(\tilde{f}(\cdot),\pi_1(\cdot))$ to satisfy the immersion conditions \eqref{immersionconditiongeneral} for the affine maps in \eqref{privacymechanism2}. Next, we design the extracting function $\pi_1^L(\cdot)$ satisfying the left invertibility condition \eqref{left_inverse}, that extracts the true global model $w^{t+1}$ from the encoded $\tilde{w}^{t+1}$. Substituting the designed immersion map \eqref{privacymechanism2} in the aggregated encoded model \eqref{distortedaggregatedModel} yields 
\begin{align}
     \tilde{w}_{a}^{t+1}&= \sum_{i=1}^{N_c} \frac{|\mathcal{D}_i|}{|\mathcal{D}|} \left(\Pi_1 w_{i,K} + b_1^t\right) \nonumber\\
     &= \Pi_1 \left(\sum_{i=1}^{N_c} \frac{|\mathcal{D}_i|}{|\mathcal{D}|} w_{i,K}\right) + b_1^t=\Pi_1 w^{t+1} + b_1^t,\label{distortedaggregatedModel2}
\end{align}
where $\tilde{w}_{a}^{t+1}$ and $w^{t+1}$ denote the encoded and original aggregated models, respectively. Given \eqref{distortedaggregatedModel2}, condition \eqref{left_inverse} can be written as
\begin{align} \label{left_inverse2}
\pi_1^L \left(\tilde{w}_{a}^{t+1} \right) =  \pi_1^L \left(\Pi_1 w^{t+1} + b_1^t \right) =  w^{t+1},
\end{align}
which trivially leads to
\begin{equation}
    \pi_1^L(\tilde{w}_{a}^{t+1}) := \Pi_1^L \tilde{w}_{a}^{t+1},\label{inversemap}
\end{equation}
since $\Pi_1^L \Pi_1=I$ and $b_1^t \in \text{ker}[\Pi_1^L]$.\\
\indent We can now state the proposed solution to Problem \ref{problem1}.
\begin{proposition}\label{proposition1}\emph{\textbf{(Solution to Problem \ref{problem1})}}
For given full rank matrix $\Pi_1 \in \mathbb{R}^{\tilde{n} \times n}$, matrix $N_1 \in \mathbb{R}^{\tilde{n} \times (\tilde{n} - n)}$ expanding the kernel of $\Pi_{1}^{L}$ \emph{(}i.e., $\Pi_{1}^{L}N_1=\mathbf{0}$\emph{)}, and random process $r_1^t \in \mathbb{R}^{(\tilde{n} - n)}$, 
the encoding map: \vspace{-1mm}
\begin{equation}\label{privacymechanisms}
 \tilde{w}^t := \Pi_1 w^t +N_1 r_1^t,
\vspace{-1mm}
\end{equation}
target optimizer:
\begin{equation}
\left\{
\begin{aligned}
&\tilde{w}_{i,0}=\tilde{w}^{t}, \\
&\tilde{w}_{i,{k+1}}= \tilde{f}(\tilde{w}_{i,k},\mathcal{D}_{i}):=\tilde{w}_{i,k} -
\Pi_1 g( \Pi^L_1 \tilde{w}_{i,k},\mathcal{D}_{i}),\label{targetoptimizersolution}\\
&k=0,1, \cdots, K-1,\\
&\tilde{w}_{i}^{t+1} = \tilde{w}_{i,K},
\end{aligned}\right.
\end{equation}
and inverse function:
\begin{equation}\label{inversefunctionsolution}
    \pi^L(\tilde{w}^{t+1}) := \Pi_1^L \tilde{w}^{t+1},
\end{equation}
provide a solution to Problem \ref{problem1}.
\end{proposition}
\emph{\textbf{Proof}}: Proposition \ref{proposition1} follows from the discussion provided in this section above.
\hfill $\blacksquare$
\subsection{Summary SIFL Algorithm solving Problem \ref{problem1}: Privacy-Preserving Aggregation}
The summary of the algorithm is as follows:
\begin{itemize}
\item \textbf{FL initialization and encoding by the server}. The server initializes the global model $w^0$ and encodes it as $\tilde{w}^0=
\Pi_1 w^0+b_1^0$. Then, it immerses the standard optimizer into the target optimization algorithm as in \eqref{targetoptimizersolution} and broadcasts $\tilde{w}^0$, the target optimizer $\Tilde{f}(\cdot)$, and other hyperparameters to clients.
\item \textbf{Local model training and update by clients}. The clients receive the current encoded global model $\tilde{w}^t$ sent by the server and update their local model parameters using their local databases $\mathcal{D}_i$ and the target optimizer \eqref{targetoptimizersolution}. Then, clients send their encoded local updates $\tilde{w}_i^{t+1}$ to the aggregator for aggregation.
\item \textbf{Global model aggregation}. The aggregator takes the average of local encoded models and sends the aggregated model $\tilde{w}_a^{t+1}$ in \eqref{distortedaggregatedModel2} to the server.
\item \textbf{Global model decoding and encoding and broadcasting by the server}. The server decodes the aggregated global model using the inverse function $\pi_1^L(\cdot)$ in \eqref{inversefunctionsolution}. Then, it encodes the new global model using the immersion map $\pi_1(\cdot)$ \eqref{privacymechanisms} and broadcasts it to all clients for the next round.
\end{itemize}
The pseudo-code of SIFL for privacy-preserving model aggregation is shown in Algorithm \ref{alg:one}.
\vspace{-2mm}
\begin{algorithm}[t]
    \caption{SIFL algorithm solving Problem \ref{problem1}: Privacy-Preserving Aggregation}\label{alg:one}
    \begin{algorithmic}
\State \textbf{Input:} {Clients $\mathcal{C}_i$ and their databases $\mathcal{D}_i$, $i \in \{1,2,...,N_c\}$, FL iterations $T$, hyperparameters of the specific gradient-descent optimizer, local iterations $K$, immersion mapping matrix $\Pi_1 \in \mathbb{R}^{\Tilde{n} \times n}$, its left inverse $\Pi_1^L$, and matrix $N_1 \in \operatorname{ker}[\Pi_1^L]$.}
\State \textbf{Output: }{Trained global model $w^T$.}\\
\textbf{Handshaking phase:}\\
The server sends target optimizer $\Tilde{f}$ as in \eqref{targetoptimizersolution}, the encoded initialized global model $\tilde{w}^0$, and other hyperparameters to clients for model update.
\For{each global iteration $t=0,1,...,T-1$}
    \State \textit{\textbf{Local Training Process by Clients:}}
    \For{each $\mathcal{C}_i$}
        \State Initialize: $\tilde{w}_{i,0} \gets \tilde{w}^{t}$.
        \For{each local iteration $k=1,2,...,(K-1)$}
        \State $\tilde{w}_{i,k+1} \leftarrow \Tilde{f}(\tilde{w}_{i,k},\mathcal{D}_{i}).$
        \State Clients send $\tilde{w}_i^{t+1}=\tilde{w}_{i,K}$ to aggregator.
        \EndFor
         \EndFor
    \State \textit{\textbf{Aggregation Process by Aggregator:}}
\State \begin{equation}
\tilde{w}_{a}^{t+1}=\sum_{i=1}^{N_c} \frac{|\mathcal{D}_i|}{|\mathcal{D}|} \tilde{w}_i^{t+1}. \nonumber
\end{equation}
\State The aggregator sends $\tilde{w}_{a}^{t+1}$ to the server.
    \State \textit{\textbf{Decoding and Encoding Process by Server}}:
     \State ${w}^{t+1}=\pi_1^L(\tilde{w}_{a}^{t+1}).$
    \For{$t<T-1$} 
    \State$\tilde{w}^{t+1}=\pi_1(w^{t+1}).$
    \EndFor
    \State Server sends encoded $\tilde{w}^{t+1}$ to clients.
    \EndFor
    \end{algorithmic}
\end{algorithm}
\begin{figure*}
  \centering
  \includegraphics[width=.95\textwidth]{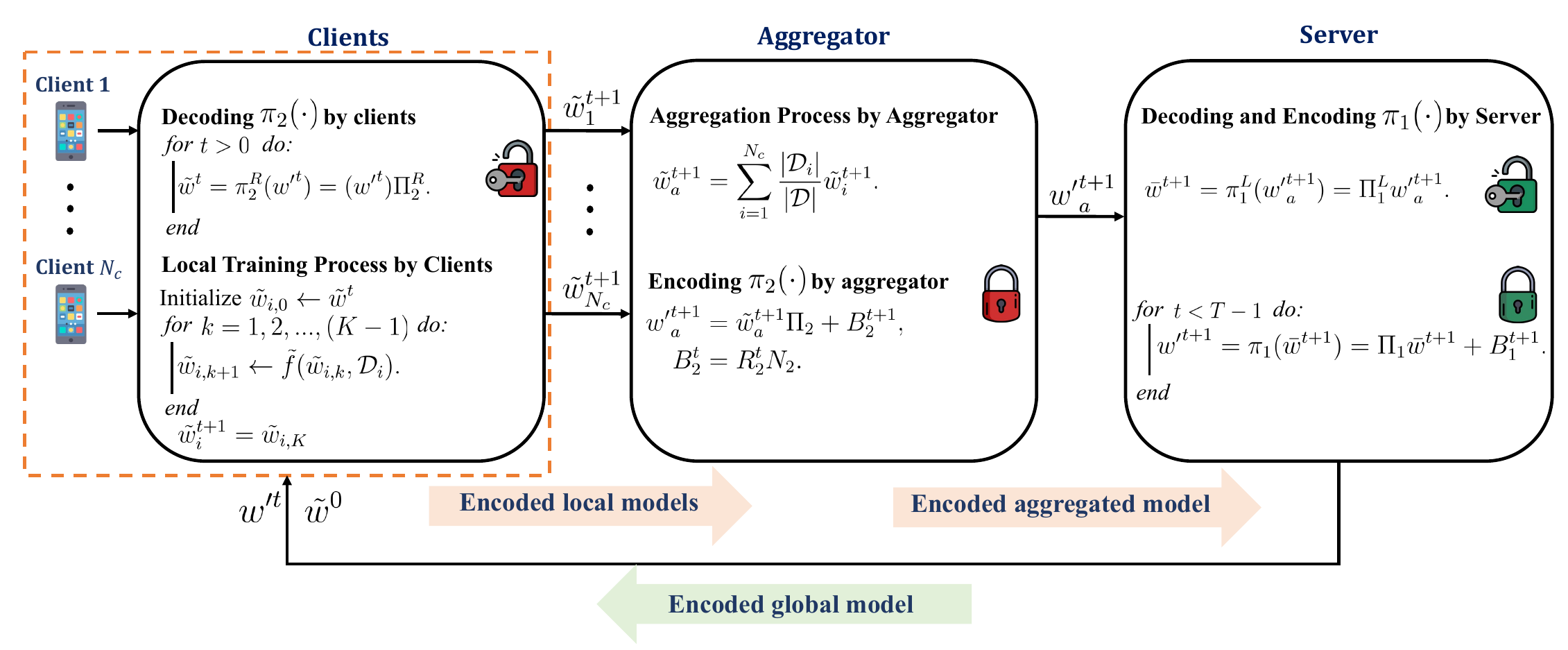}
    \caption{Flowchart of the extended SIFL method.}
    \label{flowchart}
\end{figure*}
\section{Global model privacy in FL}\label{sec4}
In the previous section, we present the immersion-based coding scheme to provide privacy for local and global models of FL. Using this scheme, none of the internal and external actors can access the original local and global models; only the server can access the exact aggregated models in each iteration. It has been shown that the server can recover local models by accessing multiple intermediate global models \cite{so2023securing}. Therefore, we next consider potential inference over the aggregated models at the server side.\\ \indent In this section, 
we provide a similar privacy-preserving mechanism based on an additional encoding of the aggregated model by the aggregator, using a mapping $\pi_2(\cdot)$, and decoding it by clients on the next iteration, to avoid server access to intermediate global models during the training process. Let $\tilde{w}^t$, $\bar{w}^t$, and ${w^{\prime}}^t$ denote the global models encoded using $\pi_1(\cdot)$ by the server, global models encoded using $\pi_2(\cdot)$ by the aggregator, and global models encoded using both $\pi_1(\cdot)$ and $\pi_2(\cdot)$, respectively. In this scheme, the aggregator encodes the aggregated model using a right invertible encoding map $\pi_2(\cdot)$ and sends the encoded aggregated model ${w^{\prime}}_a^{t+1}=\pi_2(\Tilde{w}_{a}^{t+1})$ to the server. Then, in the following iteration, clients decode the encoded aggregated model using the right inverse of the mapping $\pi_2(\cdot)$, denoted as $\pi_2^R(\cdot)$, i.e., $\pi_2 \circ \pi_2^R (s) = s$, and use the encoded model (which is still encoded by the immersion map $\pi_1(\cdot)$ of the server) for the model update. Then, the server needs to be able to apply its decoding and encoding maps, $\pi^L_1(\cdot)$ and $\pi_1(\cdot)$, to the encoded model by the aggregator ${{w}^{\prime}_a}^{t+1}=\pi_2(\Tilde{w}_{a}^{t+1})$ without decoding $\pi_2(\cdot)$. Furthermore, clients need to be able to decode $\pi_2(\cdot)$ using $\pi_2^R(\cdot)$ after encoding by the server to employ it as the initial condition for the target optimizer. Therefore, coding $\pi_2(\cdot)$ is also a homomorphic encryption scheme, meaning that the server needs to be able to apply its coding on the encoded ${{w}^{\prime}_a}^{t+1}$ without decoding $\pi_2(\cdot)$ and then, the clients need to be able to extract ${\tilde{w}}^{t+1}$ from the encoded ${w^{\prime}}^{t+1}$ using the right inverse function $\pi_2^R(\cdot)$. These conditions can be formulated as follows.\\ 
The server receives the encoded aggregated model ${{w}_a^{\prime}}^{t+1}=\pi_2({\tilde{w}}_a^{t+1})$ and decode mapping $\pi_1(\cdot)$ as follows:
\begin{equation}\label{wbar}
    \Bar{w}^{t+1}= \pi_1^L({{w}_a^{\prime}}^{t+1}).
\end{equation}
Then, server encodes $\Bar{w}^{t+1}$ using mapping $\pi_1(\cdot)$ before broadcasting it to clients as ${w^{\prime}}^{t+1}=\pi_1(\Bar{w}^{t+1})$. Clients receive ${{w}^{\prime}}^{t+1}$ and aim to retrieve ${\tilde{w}}^{t+1}$, which is still encoded using the immersion map of the server $\pi_1(\cdot)$, to employ it as the initial condition for the target optimizer. Hence, there must exist a right inverse function $\pi_2^R(\cdot)$ satisfying the following right invertibility condition:
\begin{equation}\label{rightinvertibilitycondition}
\begin{aligned}
{\tilde{w}}^{t+1}=\pi_2^R({w'}^{t+1}).
\end{aligned}
\end{equation}
Note that the aggregator needs to share the decoding map $\pi_2^R(\cdot)$ with clients in the handshaking phase. With this approach, the server does not have access to any local or intermediate global models. The problem of designing the aggregator mapping can be defined as follows.
\begin{problem}\label{problem2}\emph{\textbf{(Coding for Global Model Privacy)}} For given immersion map of the server $\pi_1(\cdot)$ \eqref{privacymechanisms} and its inverse $\pi^L_1(\cdot)$ \eqref{inversefunctionsolution}, design an immersion map $\pi_2(\cdot)$ so that: \textbf{(a)} mapping $\pi_1(\cdot)$ can be decoded after applying mapping $\pi_2(\cdot)$ to the aggregated model using $\pi_1^L(\cdot)$, i.e., \eqref{wbar}; and \textbf{(b)} there exists a function $\pi_2^R(\cdot)$ satisfying \eqref{rightinvertibilitycondition}.
\end{problem}
\subsection{Affine Solution to Problem \ref{problem2}}
\vspace{-2mm}
As we proposed in the solution to Problem \ref{problem1}, let the encoding map $\pi_2(\cdot)$ be an affine function of the form:
\begin{equation}
{w'}_a^{t+1}=\pi_2\left(\Tilde{w}_a^{t+1}\right):= \Tilde{w}_a^{t+1} \Pi_2 +B_2^{t+1}, \label{mapping2}
\end{equation}
for vector $\Pi_2 \in \mathbb{R}^{1 \times p}$, $p>1$, and matrix $B_2^t \in \mathbb{R}^{\Tilde{n} \times p}$. We design $\pi_2(\cdot)$ such that ${w'}_a^{t}=\pi_2\left(\Tilde{w}_a^{t}\right) \in \mathbb{R}^{\Tilde{n} \times p}$ is of higher dimension than $\Tilde{w}_a^{t} \in \mathbb{R}^{\Tilde{n}}$. We impose this design to create redundancy in the encoding map, which allows us to inject randomness that can be 1) traced through the coding algorithm at the server side, 2) be removed after at the clients' side, and 3) enforce an arbitrary level of DP for the aggregated models. The reason why the increase in the dimension of the aggregated model $\Tilde{w}_a^{t}$ is from the right side (rather than similar to the previous mapping $\pi_1(\cdot)$ where we increase the dimension from the left side as shown in \eqref{privacymechanism2}) is to be able to decode mapping $\pi_1(\cdot)$, with the left inverse map $\pi_1^L(\cdot)$ at the server \eqref{wbar}, and to decode mapping $\pi_2(\cdot)$ with the right inverse map $\pi_2^R(\cdot)$ at clients \eqref{rightinvertibilitycondition}.\\  
By substituting \eqref{mapping2} in the invertibility condition of mapping $\pi_1(\cdot)$ \eqref{wbar}, we have:
\begin{equation}\label{wbar2}
\Bar{w}^{t+1}= \pi_1^L( \Tilde{w}_a^{t+1} \Pi_2 +B_2^{t+1}) ={w}^{t+1} \Pi_2 + \Pi_1^L B_2^{t+1}.
\end{equation}
Hence, using the proposed structure of mapping $\pi_2(\cdot)$, mapping $\pi_1(\cdot)$ can be still decoded using $\pi_1^L(\cdot)$. Note that, to be able to apply the encoding map $\pi_1(\cdot)$ to $\Bar{w}^{t}$, according to its dimension $\Bar{w}^{t} \in \mathbb{R}^{n \times p}$, the dimension of mapping $\pi_1(\cdot)$ must be set to $\pi_1:\mathbb{R}^{n \times p} \rightarrow \mathbb{R}^{{\tilde{n}} \times p}$ (rather than the previous design in equation \eqref{privacymechanism2}, $ \pi_1:\mathbb{R}^{n} \rightarrow \mathbb{R}^{{\tilde{n}}}$). Hence, $\pi_1(\cdot)$ with this dimension is as follows:
\begin{equation}\label{newpi1}
    \pi_1(\Bar{w}^{t})=\Pi_1 \Bar{w}^{t} + B_1^t,
\end{equation}
where the additive random term $B_1^{t}$ should be designed with dimension $B_1^{t}=N_1 R_1^t\in \mathbb{R}^{\tilde{n} \times p}$, which can be achieved by setting the dimension of the random term as $R_1^t \in \mathbb{R}^{(\tilde{n}-n) \times p}$. Hence, the encoded model parameters after applying mapping $\pi_1(\cdot)$ by the server is of the following form:
\begin{equation}\label{encodedboth}
    {w^{\prime}}^{t+1}=\pi_1(\Bar{w}^{t+1})=\Pi_1 {w}^{t+1} \Pi_2 + \Pi_1 \Pi_1^L B_2^{t+1} + B_1^{t+1}.
\end{equation}
Clients receive encoded ${w^{\prime}}^{t}$ and aim to decode the aggregator coding $\pi_2(\cdot)$ using the decoding map $\pi_2^R(\cdot)$, before updating the local models by the target optimizer \eqref{targetoptimizersolution}. Substituting \eqref{encodedboth} in the right invertibility condition \eqref{rightinvertibilitycondition}, the design condition for $\pi_2^R(\cdot)$ amounts to:
\begin{equation}\label{encoding2condition}
\begin{aligned}
\pi_2^R({w'}^{t})=\pi_2^R(\Pi_1 {w}^{t} \Pi_2 + \Pi_1 \Pi_1^L B_2^{t} + B_1^{t}) = \Pi_1 {w}^{t} + b_1^{t}.
\end{aligned}
\end{equation}
Let the decoding map be of the form $ \pi_2^R({w'}^{t}) = {w'}^{t} M_3$ with $M_3 \in \mathbb{R}^{p \times 1}$. Hence, condition \eqref{encoding2condition} takes the form
\begin{equation}\label{encoding2condition2}
(\Pi_1 {w}^{t} \Pi_2 + \Pi_1 \Pi_1^L B_2^{t} + B_1^t) M_3= \Pi_1 {w}^{t} + b_1^{t}.
\end{equation}
Hence, to satisfy \eqref{encoding2condition2}, we must have $\Pi_2 M_3=I$ and $B_2^t M_3=\mathbf{0}$. Moreover, the vector $b^t_1=B_1^{t} M_3 \in \mathbb{R}^{\tilde{n} \times 1}$ should work as the additive random term in $\pi_1(\cdot)$ satisfying the necessary condition of ${b}^t_1 \in \operatorname{ker}[\Pi_1^L]$ ($\Pi_1^L {b}^t_1 =0$). It follows that: 
1) vector $M_3$ is a right inverse of $\Pi_2$, i.e.,  $M_3 = \Pi_2^R$; 2) $B_2^t$ is in the right null space of $\Pi_2^R$ ($B_2^t \in \operatorname{ker}[\Pi_2^R]$); this null space is always nontrivial because $\Pi_2$ is a row vector. Then, $b^t_1:=B_1^{t} \Pi_2^R$ always satisfy the condition $\Pi_1^L {b}^t_1 =0$ provided that $\Pi_1^L B_1^{t} =\mathbf{0}$. The final form for $\pi_2^R(\cdot)$ is given by
\begin{align}
   \pi_2^R({w'}^{t}) := {w'}^{t} \Pi_2^R.
\end{align}
At every global iteration $t$, the aggregator designs a matrix $B_2^t$ satisfying $B_2^t \Pi_2^R=0$ and uses it to construct the encoding map $\pi_2(\Tilde{w}^{t}) = \Tilde{w}^{t} \Pi_2 +B_2^t$ (the encoding scheme). The role of $B_2^t$ is to randomize the aggregated model parameters so that we can guarantee DP for intermediate global models. Using the same reasoning as for the design of $b_1^t$, to enforce that $B_2^t \in \operatorname{ker}[\Pi_2^R]$, without loss of generality, we let it be of the form $B_2^t=R_2^t N_2$ for some matrix $N_2 \in \mathbb{R}^{(p-1) \times p}$ expanding the kernel of $\Pi_2^R$ (i.e., $N_2 \Pi_2^R=\mathbf{0}$) and an arbitrary i.i.d. process $R_2^t \in \mathbb{R}^{\Tilde{n} \times (p-1)}$. This structure for $B_2^t$ always satisfies $B_2^t \Pi_2^R= R_2^t N_2 \Pi_2^R =\mathbf{0}$, for $R_2^t$ with arbitrary probability distribution.\\
\indent We can now state the proposed solution to Problem \ref{problem2}.
\begin{proposition}\label{proposition2}\emph{\textbf{(Solution to Problem \ref{problem2})}}
For given vector $\Pi_2 \in \mathbb{R}^{1 \times p}$, matrix $N_2 \in \mathbb{R}^{p-1 \times p}$ expanding the kernel of $\Pi_{2}^{R}$ \emph{(}i.e., $N_2 \Pi_{2}^{R}=\mathbf{0}$\emph{)}, and random process $R_2^t \in \mathbb{R}^{\tilde{n} \times (p-1)}$, 
the encoding map: \vspace{-1mm}
\begin{equation}\label{privacymechanisms2}
 {{w}^{\prime}}^t = \pi_2({\Tilde{w}}_a^t) := {\Tilde{w}}^t \Pi_2 +R_2^t N_2,
\vspace{-1mm}
\end{equation}
and inverse function:
\begin{equation}\label{inversefunctionsolution2}
    \pi_2^R({{w}^{\prime}}^t) := {{w}^{\prime}}^t \Pi_2^R,
\end{equation}
provide a solution to Problem \ref{problem2}.
\end{proposition}
\emph{\textbf{Proof}}: Proposition \ref{proposition2} follows from the discussion provided above.
\hfill $\blacksquare$
\begin{remark}
In Algorithm \ref{alg:one} of the method, we present a cryptography system to provide privacy for local and global models of FL in Proposition \ref{proposition1}. Using this scheme, none of the internal and external actors have access to the original local and global models, except that the server only has access to the exact aggregated model in each iteration. In the extension of the SIFL method proposed in this section, we design a similar coding to encode the aggregated model by the aggregator to avoid server access to the intermediate global models. Hence, this extension can prevent all internal actors (the server, clients, and aggregator) and external actors from accessing the intermediate local and global models of FL. 
\end{remark}
\subsection{Extended SIFL Algorithm solving Problems \ref{problem1} and \ref{problem2}: Privacy-Preserving Aggregation and Broadcasting }
The summary of the extended algorithm is as follows:
\begin{itemize}
\item \textbf{FL initialization}. The server initializes the global model $w^0$ and encodes it as $\tilde{w}^0=\Pi_1 w^0+b_1^0$. Then, it immerses the standard optimizer into the target optimization algorithm as in \eqref{targetoptimizersolution} and broadcasts $\tilde{w}^0$, the target optimizer $\Tilde{f}(\cdot)$, and other hyperparameters to clients. The aggregator sends the decoding map $\pi_2^R(\cdot)$ in \eqref{inversefunctionsolution2} to clients.
\item \textbf{Local model training and update by clients}. At the first round ($t=0$), clients use the initial model $\tilde{w}^0$ and the target optimizer \eqref{targetoptimizersolution} to train their local models. After the first round, the clients receive the encoded global model ${w^{\prime}}^t$, encoded both by the server and the aggregator. They first decode the aggregator mapping using the inverse function $\pi_2^R(\cdot)$ in \eqref{inversefunctionsolution2} to extract $\Tilde{w}^t$. Then, they update their local model parameters using their local databases $\mathcal{D}_i$ and the target optimizer \eqref{targetoptimizersolution} and send their encoded local updates $\tilde{w}_i^{t+1}$ to the aggregator for aggregation.
\item \textbf{Global model aggregation}. The aggregator takes the average of local encoded models, encodes the aggregated model $\Tilde{w}_a^{t+1}$ with $\pi_2(\cdot)$ in \eqref{privacymechanisms2}, and sends the aggregated encoded model ${{w}^{\prime}}^{t+1}$ to the server.
\item \textbf{Global model decoding, encoding, and broadcasting by the server}. The server decodes the aggregated global model using the inverse function $\pi_1^L(\cdot)$ in \eqref{inversemap}. Then, it encodes the new global model using the immersion map $\pi_1(\cdot)$ \eqref{privacymechanisms} and broadcasts it to all clients for the next round.
\end{itemize}
The pseudo-code of our extended SIFL scheme is shown in Algorithm \ref{alg:two}, and its flowchart is depicted in Figure \ref{flowchart}.
\vspace{-2mm}
\begin{algorithm}
    \caption{Extended SIFL Method: Privacy-Preserving Local and Global Models}\label{alg:two}
    \begin{algorithmic}
\State \textbf{Input: }{Set of clients $\mathcal{C}_i$ and their databases $\mathcal{D}_i$, number of FL iterations $T$, hyperparameters of the specific gradient descent optimizer, number of local optimizer iterations $K$, immersion mapping matrix $\Pi_1 \in \mathbb{R}^{\Tilde{n} \times n}$, its left inverse $\Pi_1^L$, matrix $N_1 \in \operatorname{ker}[\Pi_1^L]$, vector $\Pi_2\in \mathbb{R}^{1 \times p}$, its right inverse $\Pi_2^R$, matrix $N_2 \in \operatorname{ker}[\Pi_2^R]$.}
\State \textbf{Output: }{Trained global model $w^T$.}
\State \textbf{Handshaking phase:}
\State The server sends target optimizer $\Tilde{f}(\cdot)$ as in \eqref{targetoptimizersolution}, the encoded initialized global model $\tilde{w}^0$, and other hyperparameters to clients for model update. The aggregator sends the decoding key $\Pi_2^R$ to clients.
\For{each global iteration $t=0,1,...,T-1$}
    \For{each $\mathcal{C}_i$}
        \State \textbf{Decoding $\pi_2(\cdot)$ by Clients:}\\
         \For{$t>0$}
         \State \begin{equation}
            \tilde{w}^{t}=\pi^R_2({w'}^{t})= ({w'}^{t}) \Pi_2^R.  
             \end{equation}
         \EndFor
         \State\textit{\textbf{Local Training Process by Clients:}}
        \State Initialize: $\tilde{w}_{i,0} \gets \tilde{w}^{t}$
        \For{each local iteration $k=1,2,...,(K-1)$}
        \State \begin{equation}
            \tilde{w}_{i,k+1} \leftarrow \tilde{f}(\tilde{w}_{i,k} ,\mathcal{D}_{i}).\nonumber
        \end{equation}\EndFor
        \State $\mathcal{C}_i$ sends $\tilde{w}^t_{i}$ to the aggregator.\EndFor    
    \State \textit{\textbf{Aggregation Process by Aggregator:}}
    \State \begin{equation}
    \tilde{w}_a^{t+1}=\sum_{i=1}^{N_c} \frac{|\mathcal{D}_i|}{|\mathcal{D}|} \tilde{w}^{t+1}_{i}.\nonumber
    \end{equation}
   \State {\textbf{Encoding $\pi_2(\cdot)$ by Aggregator:}}
   \State \begin{equation}
   \begin{aligned}
       {w'}_a^{t+1} &= \tilde{w}_a^{t+1} \Pi_2 + B_2^{t+1},\\
       B_2^t&= R_2^t N_2.\nonumber
   \end{aligned}
   \end{equation}
   \State The aggregator sends ${w'}_a^{t+1}$ to the server.
\State \textit{\textbf{Decoding and encoding $\pi_1(\cdot)$ by Server:}}
     \State$\bar{w}^{t+1}=\pi_1^L({w'}_a^{t+1})= \Pi_1^L {w'}_a^{t+1}.$
     \For{$t<T-1$}
    \State \begin{equation}
         {w'}^{t+1}=\pi_1(\bar{w}^{t+1})=\Pi_1 \bar{w}^{t+1}+B_1^{t+1}.\nonumber
    \end{equation} 
    \EndFor
    \State Server sends encoded ${w'}^{t+1}$ to clients.\EndFor
    \end{algorithmic}
\end{algorithm}
\section{Privacy Guarantees}\label{sec5}
\vspace{-2mm}
The private element we consider in the proposed scheme is privacy of the clients' local databases $\mathcal{D}_i$. In what follows, we focus on how to enforce differential privacy of the encoding mechanisms $\pi_1(\cdot)$ and $\pi_2(\cdot)$ by properly selecting the random processes $R_1^t$ and $R_2^t$ and the encoding matrices in the affine maps of the server $\tilde{w}^t = \Pi_1 w^t + N_1 R^t_1$ and the aggregator $w^{\prime}_t = \tilde{w}^t \Pi_2 + R^t_2 N_2$.\\
\indent We provide a tailored solution to guarantee DP for the class of mechanisms that we consider in \eqref{privacymechanisms} and \eqref{privacymechanisms2}. In particular, we prove that the proposed scheme, with full-column rank encoding $\Pi_1$ and vector $\Pi_2$, can provide any desired level of differential privacy without reducing the accuracy and performance of the original algorithm.
\vspace{-1mm}
\subsection{Differential Privacy}
\vspace{-2mm}
In the context of databases, $(\epsilon, \delta)$-Differential Privacy (DP) \cite{Dwork} was introduced as a probabilistic framework to quantify privacy of probabilistic maps. The constant $\epsilon \geq 0$ quantifies how similar (different) are outputs of a mechanism on \emph{adjacent} datasets, say $\mathcal{D}$ and $\mathcal{D}^{\prime}$, and $\delta$ is a constant shift used when the ratio of the probabilities of $\mathcal{D}$ and $\mathcal{D}^{\prime}$ under the mechanism cannot be bounded by $e^\epsilon$ (see Definition \ref{definition2} below). With an arbitrarily given $\delta$, a mechanism with a smaller $\epsilon$ makes \emph{adjacent} databases, $(\mathcal{D}$ and $\mathcal{D}^{\prime})$, less distinguishable and hence more private.
\begin{definition}[Adjacency \cite{Dwork}]\label{definition1}\emph{:}
Let $\mathcal{X}$ denote the space of all possible datasets. We say that $\mathcal{D} \in \mathcal{X}$ and $\mathcal{D}^{\prime} \in \mathcal{X}$ are adjacent if they differ on a single element.
\end{definition}
\begin{definition}[$(\epsilon,\delta)$-Differential Privacy \cite{Dwork}]\label{definition2}\emph{:} The random mechanism $\mathcal{M}: \mathcal{X} \rightarrow \mathcal{R}$ with domain $\mathcal{X}$ and range $\mathcal{R}$ is said to provide $(\epsilon,\delta)$-differential privacy, if for any two adjacent datasets $\mathcal{D}, \mathcal{D}^{\prime} \in \mathcal{X}$ and for all measurable sets $\mathcal{S} \subseteq \mathcal{R}$\emph{:}
\begin{equation}\label{DP}
\operatorname{Pr}(\mathcal{M}(\mathcal{D}) \in  \mathcal{S}) \leq e^\epsilon \operatorname{Pr}\left(\mathcal{M}\left(\mathcal{D}^{\prime}\right) \in  \mathcal{S}\right)+\delta.
\end{equation}
\end{definition}
\vspace{-1mm}
If $\delta=0$, $\mathcal{M}$ is said to satisfy $\epsilon$-differential privacy. From Definition \ref{definition2}, we have that a mechanism provides DP if its probability distribution satisfies \eqref{DP} for some $\epsilon$ and $\delta$. Then, if we seek to design the mechanism to guarantee DP, we need to shape its probability distribution. This is usually done by injecting noise into the data we seek to encode. The noise statistics must be designed in terms of the sensitivity of the data to be encoded. Sensitivity refers to the maximum change of the data due to the difference in a single element of the dataset \cite{Dwork}.
\begin{definition} [Sensitivity]\label{defSensitivity}\emph{:} Given adjacent datasets $\mathcal{D}, \mathcal{D}^{\prime} \in \mathcal{X}$, and a query function $q: \mathcal{X} \rightarrow \mathcal{R}$ (a deterministic function of datasets) where the output space $\mathcal{R}$ is equipped with a norm denoted $\left\|\cdot \right\|_R$, the sensitivity of $q(\cdot)$ is formulated as $\Delta^q_R=\sup _{\mathcal{D}, \mathcal{D}^{\prime}}\left\|q\right(\mathcal{D} \left) - q\left(\mathcal{D}^{\prime} \right)\right\|_R$.
\end{definition}
The differential privacy mechanism $\mathcal{M}$ must be designed to ensure that the DP condition \eqref{DP} holds. According to Definition \ref{defSensitivity}, the sensitivity of the query to which this mechanism is applied determines the design of its variables.
\vspace{-1mm}
\subsection{Immersion-based Coding Differential Privacy Guarantee}
\vspace{-2mm}
We concretely formulate the problem of designing the variables of privacy coding mechanisms $\pi_1(\cdot)$ and $\pi_2(\cdot)$ in \eqref{newpi1} and \eqref{privacymechanisms2} that, at each global iteration, guarantee the privacy of local databases. 
We wish to design matrices $\Pi_1$, $\Pi_2$, and random variables $R_1^t$ and $R_2^t$ such that the privacy mechanisms $\pi_1(\cdot)$ and $\pi_2(\cdot)$ for distorting $\tilde{w}^t_i$ and ${w^{\prime}}^t$ are $(\tilde{\epsilon}, \tilde{\delta})$ and $(\epsilon^{\prime},\delta^{\prime})$-Differentially private.
\begin{problem}\label{problem3}\emph{\textbf{(Element-Wise Differential Privacy)}} Given a sequence of desired privacy levels $(\tilde{\epsilon}, \tilde{\delta})$ and $(\epsilon^{\prime},\delta^{\prime})$, design the variables of privacy mechanisms in \eqref{newpi1} and \eqref{privacymechanisms2} such that at global iteration $t$, each element of vector $\tilde{w}^t_i$, $\tilde{w}^t_{i,j}$, and each element of matrix ${w^{\prime}}^t$, ${w_{j,m}^{\prime\,t}}$, $j \in \{1,...,\tilde{n}\}$ and $m \in \{1,...,p\}$, are $(\tilde{\epsilon}, \tilde{\delta})$ and $(\epsilon^{\prime},\delta^{\prime})$-Differentially private, respectively, for any measurable $\mathcal{S} \subset \mathbb{R}$, i.e.,
\begin{equation}\label{dpcondition}
    \left\{\begin{array}{l}\begin{aligned}
\mathbb{P}\left( \tilde{w}^t_{i,j}(\mathcal{D}_i)\in \mathcal{S}\right) &\leq e^{\tilde{\epsilon}} \mathbb{P}\left(\tilde{w}^t_{i,j}(\mathcal{D}^{\prime}_i)\in \mathcal{S} \right)+\tilde{\delta},\\
\mathbb{P}\left( {w_{j,m}^{\prime\,t}}(\mathcal{D}_i) \in \mathcal{S}\right) &\leq e^{\epsilon^{\prime}} \mathbb{P}\left({w_{j,m}^{\prime\,t}}(\mathcal{D}^{\prime}_i)\in \mathcal{S}\right)+\delta^{\prime}, \\
\text { for adjacent }\left(\mathcal{D}_i, \mathcal{D}^{\prime}_i\right).\end{aligned}  
    \end{array}\right.
\end{equation}
\end{problem}
\vspace{-1mm}
\subsection{Solution to Problem \ref{problem3}}
\vspace{-2mm}
As standard in DP, we consider two cases for stochastic processes $R_1^t$ and $R_2^t$, Laplace and Gaussian distributions, and prove DP guarantees for both scenarios.\\
Starting with the Laplace additive noise scenario, let the independent stochastic process $R_1^t$ and $R_2^t$ follow multivariate i.i.d. Laplace distributions with means $E[R_1^t] =: \mu_1 \in \Real^{(\tilde{n} - n) \times p}$ and $E[R_2^t] =: \mu_2 \in \Real^{\tilde{n} \times (p-1)}$, and covariance matrices $E[(R_1^t-\mu_1)(R_1^t-\mu_1)^\top] =: \sigma_1 I_{(\tilde{n} - n)}$ and $E[(R_2^t-\mu_2)(R_2^t-\mu_2)^\top] =: \sigma_2 I_{\tilde{n}}$, for some $\sigma_1, \sigma_2>0$, i.e., $R_1^t \sim \operatorname{Laplace }(\mu_1, \sigma_1 I_{(\tilde{n} - n)})$ and $R_2^t \sim \operatorname{Laplace }(\mu_2, \sigma_2 I_{\tilde{n}})$.\\
\indent We start with the privacy guarantee for ${w}^t_i$. According to Definition \ref{defSensitivity}, given adjacent local databases $\mathcal{D}_i, \mathcal{D}^{\prime}_i \in \mathcal{X}_i$, where $\mathcal{X}_i$ denotes the space of all user data sets, the sensitivity of ${w}^t_i$ defined is as follows:
\begin{equation}\label{sensitivityy}
    \Delta^{w_i}_1=\sup _{\mathcal{D}_i, \mathcal{D}^{\prime}_i}\left\| {w}^t_i \right(\mathcal{D}_i \left) - {w}^t_i\left(\mathcal{D}^{\prime}_i \right)\right\|_1.
\end{equation}
For simplicity, in what follows, we write ${w}^t_i(\mathcal{D}_i)$ and ${w}^t_i(\mathcal{D}^{\prime}_i)$ as ${w}^t_i$ and ${{w}^t_i}^{\prime}$. Because $R_1^t \sim \operatorname{Laplace }(\mu_1, \sigma_1 I)$, and given the privacy encoding mechanisms $\tilde{w}^t_i= \Pi_1 {w}^t_i + b_1^t$, with $b_1^t=N_1 R_1^t \Pi_2^R$ in the extended SIFL, each element of $\tilde{w}^t_i$ also follows a Laplace distribution:
\begin{equation}
\tilde{w}^t_{i,j} \sim \operatorname{Laplace}\left(\Pi_{1}^{j} {w}^t_{i} + N_1^{j} \mu_1 \Pi_2^R,||N_1^{j}||_2 \sigma_1 ||\Pi_2^R||_2\right),
\end{equation}
where $\tilde{w}^t_{i,j}$ is the $j^{th}$ element of $\tilde{w}^t_i$, and $\Pi_1^{j}$ and $N_1^{j}$ are the $j^{th}$ rows of $\Pi_1$ and $N_1$, respectively. It follows that:
\begin{equation}\label{eqproof}
\begin{aligned}
    &\mathbb{P}\left(\tilde{w}^t_{i,j}(\mathcal{D}_i) \in {\mathcal{S}} \right)\\ &\hspace{2mm}=\left(\frac{1}{2  ||N_1^{j}||_2 \sigma_1 ||\Pi_2^R||_2}\right) \int_{\mathcal{S}}  e^{\frac{-\left\|p-(\Pi_{1}^{j} {w}^t_{i} +N_1^{j} \mu_1 \Pi_2^R)\right\|_{1}}{ ||N_1^{j}||_2 \sigma_1 ||\Pi_2^R||_2}} d p\\[1mm] &\hspace{2mm}
    \stackrel{(\mathrm{a})}{\leq} \frac{ e^{\frac{\left\|\Pi_{1}^{j}\left({w}^t_i - {{w}^t_i}^{\prime}\right)\right\|_1}{ ||N_1^{j}||_2 \sigma_1 ||\Pi_2^R||_2}}}{2 ||N_1^{j}||_2 \sigma_1 ||\Pi_2^R||_2} \int_{\mathcal{S}} e^{\frac{-\left\|p-(\Pi_{1}^{j} {{w}^t_i}^{\prime} + N_1^{i} \mu_1 \Pi_2^R)\right\|_{1}}{ ||N_1^{j}||_2 \sigma_1 ||\Pi_2^R||_2}}d p \\[1mm] &\hspace{2mm}= e^{\frac{\left\|\Pi_{1}^{j}\left({w}^t_i  - {{w}^t_i}^{\prime}\right)\right\|_1}{||N_1^{j}||_2 \sigma_1 ||\Pi_2^R||_2}}\quad \mathbb{P}\left(\tilde{w}^t_{i,j}(\mathcal{D}^{\prime}_i) \in \mathcal{S} \right),
    \end{aligned}
\end{equation}
where inequality (a) follows from the triangle inequality $ -\left\|p -(\Pi_{1}^{j} {w}^t_i +N_1^{j} \mu_1 \Pi_2^R)\right\|_{1} \leq \left\|\Pi_{1}^{j}\left({w}^t_i- {{w}^t_i}^{\prime}\right)\right\|_1
    -\left\|p-(\Pi_{1}^{j} {{w}^t_i}^{\prime} +N_1^{j} \mu_1 \Pi_2^R)\right\|_{1}$.\\
Due to the sensitivity relation \eqref{sensitivityy}, we have
\begin{equation}\label{equp}
\left\|\Pi_{1}^{j}\left({w}^t_i - {{w}^t_i}^{\prime}\right) \right\|_1 \leq \left\|\Pi_{1}^{j}\right\|_1 \Delta^{w_i}_1.
\end{equation}
Hence, substituting \eqref{equp} in \eqref{eqproof} implies
\begin{equation}
    \mathbb{P}\left(\tilde{w}^t_{i,j}(\mathcal{D}_i) \in {\mathcal{S}} \right) \le e^{\frac{\left\|\Pi_{1}^{j}\right\|_1 \Delta^{w_i}_1}{ ||N_1^{j}||_2 \sigma_1 ||\Pi_2^R||_2}}\quad \mathbb{P}\left(\tilde{w}^t_{i,j}(\mathcal{D}^{\prime}_i) \in {\mathcal{S}} \right).
\end{equation}
Therefore, $\epsilon^{\Tilde{w}}$-differential privacy (with $\delta^{\Tilde{w}}=0$) of each local model parameter $\tilde{w}^t_{i,j}$ for all $ j \in \{1,2,\dots,\tilde{n}\}$, is guaranteed for $\Pi_1$, $N_1$, $\Pi_2^R$, and $\sigma_1$ satisfying:
\begin{equation}\label{dpnoisewt}
    \frac{\left\|\Pi_{1}^{j}\right\|_1 \Delta^{w_i}_1}{ ||N_1^{j}||_2 \sigma_1 ||\Pi_2^R||_2} \le \epsilon^{\Tilde{w}}.
\end{equation}
Following the same steps, it can be shown that each element of the encoded global model ${w^{\prime}}^t = \Pi_1 {w}^t \Pi_2 + B_1^t + \Pi_1 \Pi_1^L B^t_2$, ${w^{\prime}}^t _{j,m}$, with $B_1^t= N_1 R_1^t$ and  $B_2^t=R_2^t N_2$ is $\epsilon^{\prime}$-Differentially private in the sense of\eqref{dpcondition}) for $\Pi_1$, $\Pi_2$, $N_1$, $N_2$, $\sigma_1$, and $\sigma_2$ satisfying:
\begin{equation}\label{dpnoisewp}
    \frac{\left\|\Pi_{1}^{j}\right\|_1 \Delta^{w}_1 \left\|\Pi_{2}^{m}\right\|_1 }{ ||N_1^{j}||_2 \sigma_1+ ||\Pi^j_1||_2 ||\Pi^L_1||_2 ||N_2||_2 \sigma_2} \le \epsilon^{\prime},
\end{equation}
for all $j \in \{1,2,\dots,\tilde{n}\}$ and $m \in \{1,2,\dots,p\}$, where $\Pi_{1}^{j}$ and $N_1^{j}$ are the $j^{th}$ rows of matrices $\Pi_1$ and $N_1$, respectively, $\Pi_{2}^{m}$ is the $m^{th}$ element of vector $\Pi_{2}$, $N_2^{m}$ is $m^{th}$ column of $N_2$, and $\Delta^{w}_1$ is the sensitivity of the global model given by $\Delta^{w}_1=\sup _{\mathcal{D}_i, \mathcal{D}^{\prime}_i}\left\| {w}^t \right(\mathcal{D}_i \left) - {w}^t\left(\mathcal{D}^{\prime}_i \right)\right\|_1$. %
Hence, according to the differential privacy conditions, \eqref{dpnoisewt} and \eqref{dpnoisewp}, to improve privacy guarantees by decreasing $\epsilon^{\Tilde{w}}$ and $\epsilon^{w^{\prime}}$, we need to design $\Pi_1$ and $\Pi_2$ as small as possible, while designing $\sigma_1$ and $\sigma_2$, the standard deviations of $R^t_1$ and $R^t_2$, and $||N_1^{i}||_2$ and $||N_2^{j}||_2$ as large as possible. From \eqref{newpi1} and \eqref{privacymechanisms2}, it is obvious that by choosing small $\Pi_1$ and $\Pi_2$, and large $\sigma_1$ and $\sigma_2$, $\tilde{w}_i^t$ and ${w^{\prime}}^t$ are close to additive random terms $b^t_1$ and $B_1^t + \Pi_1 \Pi_1^L B^t_2$, and practically independent from ${w}_i^t$ and ${w}^t$, respectively.\\ 
\indent Note that $||N_1^{i}||_2$, $i \in \{1,\dots,\tilde{n}_y \}$ and $||N_2^{j}||_2$, $j \in \{1,\dots,p \}$, must be nonzero, i.e., we need to design $N_1$ without zero rows and $N_2$ without zero columns. The latter is not a technical constraint as, for a given $N_1$ and $N_2$ with nonzero rows and columns, respectively, $\Pi_1$ and $\Pi_2$ can be obtained by solving the equations $\Pi_1^L N_1=\mathbf{0}$ and $N_2 \Pi_2^R=\mathbf{0}$ and computing the right inverse of $\Pi_1^L$ and left inverse of $\Pi_2^R$, respectively.\\
The conditions on $\Pi_1$, $N_1$, and $\sigma_1$ to guarantee a desired level of privacy are provided in the following theorem.
\begin{theorem}\label{theoremLaplace}\emph{\textbf{(Differential Privacy through Laplace additive noises)}}
Consider given Laplace processes $R_1^t \sim \operatorname{Laplace }(\mu_1, \sigma_1 I_{(\tilde{n} - n)})$ and $R_2^t \sim \operatorname{Laplace }(\mu_2, \sigma_2 I_{\tilde{n}})$ with standard deviation $\sigma_1$ and $\sigma_2$, respectively, full-rank matrix $\Pi_1 \in \mathbb{R}^{\tilde{n} \times n}$, $\Pi_2 \in \mathbb{R}^{1 \times p}$, matrices $N_1 \in \mathbb{R}^{\tilde{n} \times (\tilde{n} - n)}$ and $N_2 \in \mathbb{R}^{(p-1)\times p}$ expanding the kernels of $\Pi_{1}^{L}$ and $\Pi_2^{R}$, respectively, that satisfy the following conditions: 
\begin{equation}\label{ineq laplace all}
\left\{ \begin{aligned}
&\frac{\left\|\Pi_{1}^{j}\right\|_1 \Delta^{w_i}_1}{ ||N_1^{j}||_2 \sigma_1 ||\Pi_2^R||_2} \le \epsilon^{\Tilde{w}},\\
&\frac{\left\|\Pi_{1}^{j}\right\|_1 \Delta^{w}_1 \left\|\Pi_{2}^{m} \right\|_1 }{ ||N_1^{j}||_2 \sigma_1+ ||\Pi^j_1||_2 ||\Pi^L_1||_2 ||N_2||_2 \sigma_2} \le \epsilon^{\prime},
\end{aligned}\right.
\end{equation}
$\tilde{\epsilon}$ and $\epsilon^{\prime}$-differential privacy guarantee are warranted for each element of vector $\Tilde{w}^t_i$, $\Tilde{w}^t_{i,j}$, and each element of matrix ${w^{\prime}}^t$, ${w_{j,m}^{\prime\,t}}$, respectively, for $j \in \{1,...,\Tilde{n} \}$ and $m \in \{1,...,p \}$.
\end{theorem}
\emph{\textbf{Proof}}: The proof follows from the discussion provided in this section above.
\hfill $\blacksquare$\\
\indent The differential privacy guarantees are also provided when the additive noises in the privacy mechanisms \eqref{newpi1} and \eqref{privacymechanisms2} are Gaussian. Let the independent stochastic processes $R_1^t$ and $R_2^t$ following multivariate i.i.d. Gaussian distributions with $E[R_1^t] =\mathbf{0}$ and $E[R_2^t]  =\mathbf{0}$, and covariance matrices $E[(R_1^t)(R_1^t)^\top] =: \sigma_1 I_{(\tilde{n} - n)}$ and $E[(R_2^t)(R_2^t)^\top] =: \sigma_2 I_{\tilde{n}}$, for some $\sigma_1, \sigma_2>0$, i.e., $R_1^t\sim \mathcal{N}(\mathbf{0}, \sigma_1 I_{(\tilde{n} - n)})$ and $R_2^t \sim \mathcal{N}(\mathbf{0}, \sigma_2 I_{\tilde{n}})$. In this case, the conditions on $\Pi_1$, $N_1$, $\Pi_2^R$, $\sigma_1$, and $\sigma_2$ to guarantee a desired level of privacy are provided in the following theorem.
\begin{theorem}\label{theoremGaussian}\emph{\textbf{(Differential Privacy through Gaussian additive noises)}}
Consider Gaussian processes $R_1^t\sim \mathcal{N}(\mathbf{0}, \sigma_1 I_{(\tilde{n} - n)})$ and $R_2^t \sim \mathcal{N}(\mathbf{0}, \sigma_2 I_{\tilde{n}})$ with standard deviation of Gaussian noises $\sigma_1$ and $\sigma_2$, full rank matrix $\Pi_1 \in \mathbb{R}^{\tilde{n} \times n}$, $\Pi_2 \in \mathbb{R}^{1 \times p}$, matrices $N_1 \in \mathbb{R}^{\tilde{n} \times (\tilde{n} - n)}$ and $N_2 \in \mathbb{R}^{(p-1)\times p}$ expanding the kernels of $\Pi_{1}^{L}$ and $\Pi_2^{R}$, respectively, satisfying the following conditions: 
\begin{equation}\label{ineq gauss3}
\left\{ \begin{aligned}
&(||N_1^{j}||_2 \sigma_1 ||\Pi_2^R||_2)^2  -\frac{||\Pi_{1}^{j}||_2^2(\Delta^{w_i}_2)^2
}{2 \tilde{\epsilon}}\\
&\quad\quad - \frac{||\Pi_{1}^{j}||_2\Delta^{w_i}_2 
}{\tilde{\epsilon}} Q^{-1}(\tilde{\delta})(||N_1^{j}||_2 \sigma_1 ||\Pi_2^R||_2)\geq 0,\\
&(||N_1^{j}||_2 \sigma_1 +||N_2^{m}||_2 \sigma_2)^2 -\frac{||\Pi_{1}^{j}||_2^2(\Delta^w_2)^2 ||\Pi_{2}^{m}||^2_2
}{2 \tilde{\epsilon}}\\
&\quad\quad \quad\quad - \frac{||\Pi_{1}^{j}||_2\Delta^w_2
||\Pi_{2}^{m}||_2}{\tilde{\epsilon}} Q^{-1}(\tilde{\delta})(||N_1^{j}||_2 \sigma_1 \\&\quad\quad \quad\quad+||\Pi^j_1||_2 ||\Pi^L_1||_2 ||N_2||_2 \sigma_2) \geq 0,
\end{aligned}\right.
\end{equation}
with the $Q$-function $Q(x):=\frac{1}{\sqrt{2 \pi}} \int_x^{\infty} e^{-u^2 / 2} d u$, i.e., the tail distribution of the standard normal distribution. Then, $(\tilde{\epsilon}, \tilde{\delta})$ and $(\epsilon^{\prime},\delta^{\prime})$- Differential Privacy guarantees are warranted for each element of $\Tilde{w}^t_i$, $\Tilde{w}^t_{i,j}$, and each element of ${w^{\prime}}^t$, ${w_{j,m}^{\prime\,t}}$, respectively, for $j \in \{1,...,\Tilde{n} \}$ and $m \in \{1,...,p \}$, at global iteration $t$ of the FL algorithm.
\end{theorem}
\vspace{-1mm}
\emph{\textbf{Proof:}} See Appendix \ref{Prooftheoremgaussian}.\hfill $\blacksquare$\\
The inequality condition in \eqref{ineq gauss3} shows the relation between privacy levels, ($\tilde{\epsilon},\tilde{\delta}$) and (${\epsilon}^{\prime},{\delta}^{\prime}$), and privacy mechanism design variables ($\sigma_1,\sigma_2, \Pi_1, \Pi_2$, $N_1$). To obtain a higher level of privacy, which can be achieved by reducing the amount of ($\tilde{\epsilon},\tilde{\delta}$) and (${\epsilon}^{\prime},{\delta}^{\prime}$), we need to choose smaller $\Pi_1$ and $\Pi_2$, larger $N_1$, and larger noise standard deviations $\sigma_1$ and $\sigma_2$.
%
\begin{remark}\label{remark3}
The sizes of the additive noises required to guarantee a desired level of DP for local and global FL models are multiples of the sensitivities of local databases for local and global models, $\Delta^{w_i}_l$ and $\Delta^{w}_l$ for $l={1,2}$. These sensitivities can be calculated by $\Delta^{w_i}_l =\frac{2C}{|\mathcal{D}_i|}$ and $\Delta^{w}_l =\frac{2C}{|\mathcal{D}|}$ where $C$ is a clipping threshold for bounding $w_i$ \cite{wei2020federated}. Since in the SIFL method the distortion induced by these noises can be removed by the server and clients, the noises do not need to be small. Therefore, we can choose a large clipping threshold to avoid distorting the FL performance.   
\end{remark}
\begin{remark}
In \eqref{dpnoisewt}, \eqref{dpnoisewp}, and Theorem \ref{theoremGaussian}, we propose the design conditions for the variables of the privacy mechanisms, $\Pi_1$, $\Pi_2$, $\sigma_1$, and $\sigma_2$, to guarantee $(\tilde{\epsilon}, \tilde{\delta})$ and $(\epsilon^{\prime},\delta^{\prime})$- DP for local and global models. It has been shown that a differentially private algorithm is perfectly secret if the set of differential privacy levels is reached zero \cite{hayati2024privacy}. Hence, to have perfect secrecy for coding mechanisms in SIFL methods, we need $(\tilde{\epsilon}, \tilde{\delta}, {\epsilon}^{\prime}, {\delta}^{\prime}) \rightarrow 0$. Although this cannot be achieved due to the numerical limits, $(\tilde{\epsilon}, \tilde{\delta}, {\epsilon}^{\prime}, {\delta}^{\prime})$ can be arbitrarily small by selecting small $\Pi_1$ and $\Pi_2$ and large $\sigma_1$ and $\sigma_2$, respectively.
\end{remark}
\vspace{-1mm}
\section{Simulation Experiments}\label{sec6}
\vspace{-2mm}
\begin{figure*}[t]
\centering
\subfloat{\includegraphics[width = 1.75in]{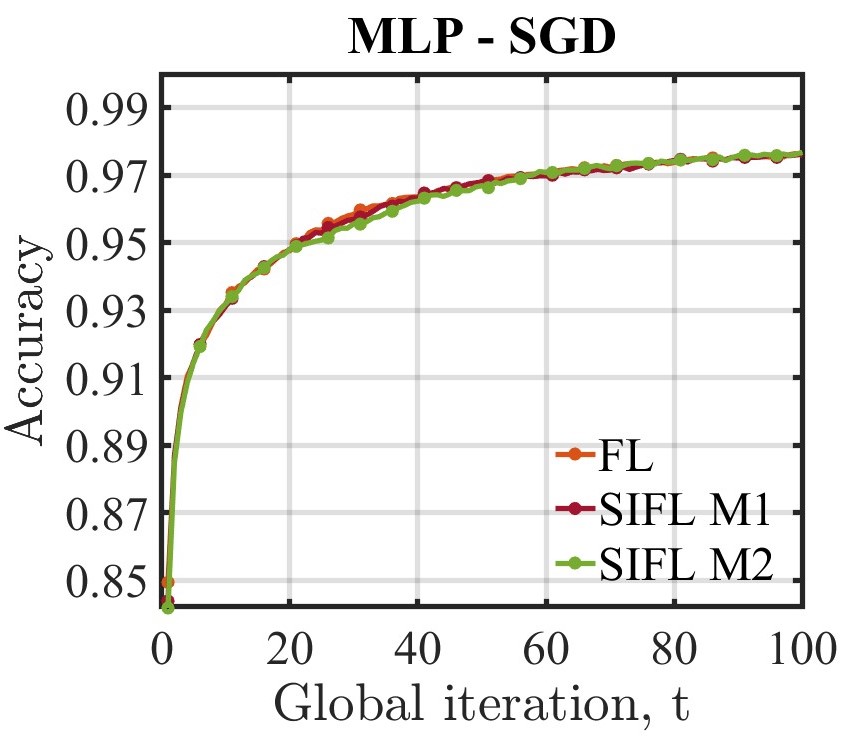}} 
\subfloat{\includegraphics[width = 1.75in]{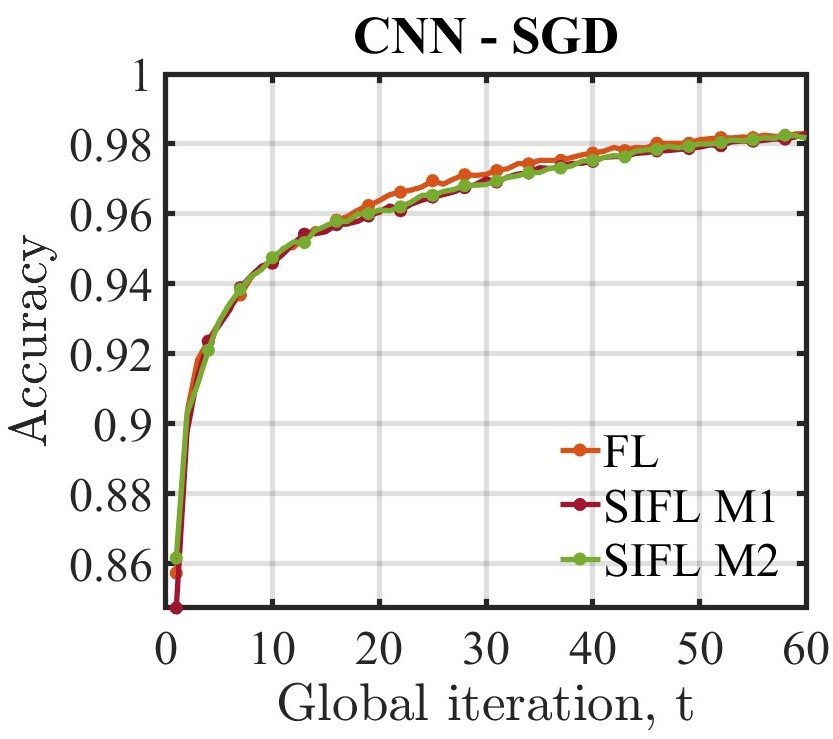}} 
\subfloat{\includegraphics[width = 1.75in]{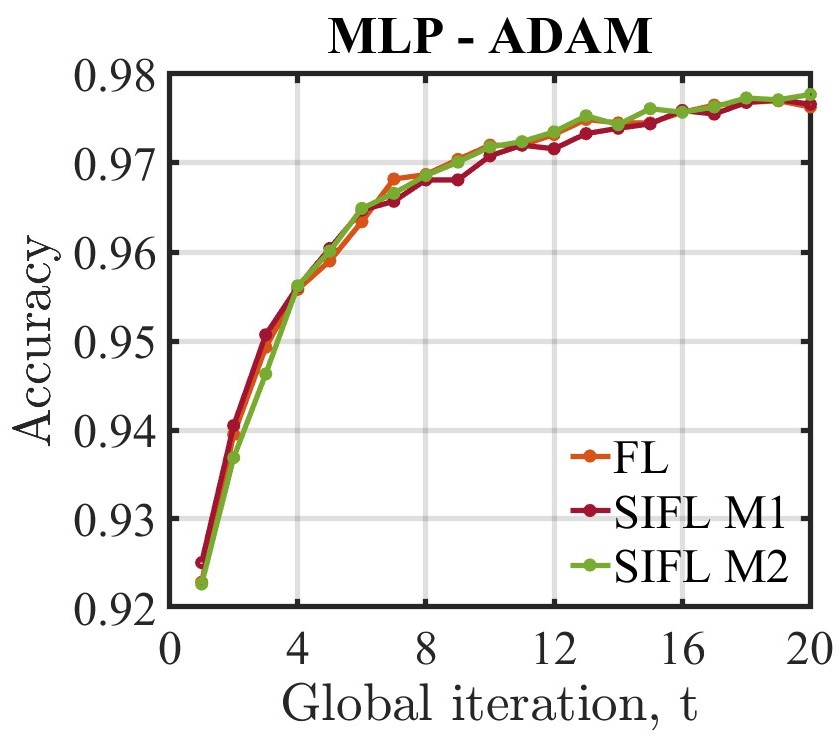}} 
\subfloat{\includegraphics[width = 1.75in]{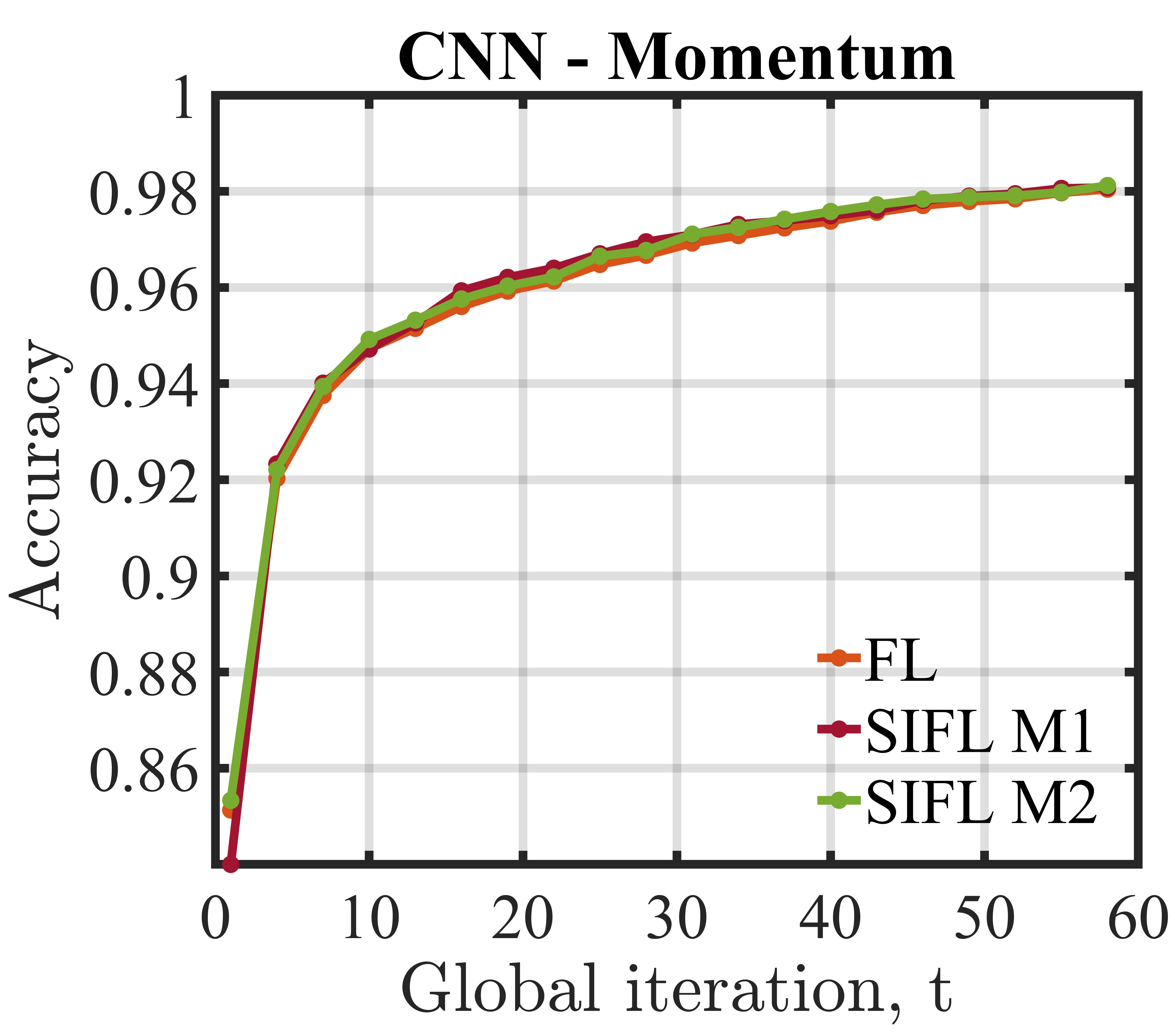}} 
\caption{The comparison of the accuracy of FL network in each iteration with and without the proposed privacy mechanism.}
\label{fig:accuracy}
\end{figure*}
\subsection{Experimental Setup}
We examine the experimental results for both SIFL methods, on three neural network models, namely the Multi-Layer Perceptron (MLP) and two different Convolutional Neural Networks (CNN). Our investigation involves utilizing different optimization tools \cite{ruder2016overview}, e.g., Adam, SGD, and Momentum, on the MNIST \cite{lecun1998gradient} and Fashion-MNIST \cite{xiao2017fashion} databases. The experimental details are described as follows:
\begin{itemize}
\item{Datasets:} We test our algorithm on the MNIST database for handwritten digit recognition and Fashion-MNIST database for Zalando's article images classification, both containing 60000 training and 10000 testing instances of 28× 28 size gray-level images and 10 classes.
\item{Models:} The MLP model is a feed-forward deep neural network with ReLU units and a softmax layer of 10 classes (corresponding to the ten digits) with two hidden layers containing 200 hidden units containing 199,210 parameters. The first CNN model consists of two $3 \times 3$ convolutional layers followed by the $2 \times 2$ max pooling layer and ReLu activation function. The first layer has 32 channels, while the second has 64 channels. The fully connected layer has 128 units and takes the flatted output of the second convolutional layer as the input. The CNN model contains $1,199,882$ parameters. The second CNN model (CNN2) has two $5 \times 5$ convolution layers (the first with 32 channels, the second with 64, each followed with $2 \times 2$ max pooling), a fully connected layer with 512 units and ReLu activation, and a final softmax output layer, containing $582,026$ weight parameters. Cross-entropy is employed as the loss function in all three networks.
\item{Optimization tools:} As optimization algorithms, the SGD, Momentum, and Adam optimizers with learning rates 0.01, 0.01, and 0.001, and local epoch $K=2$, $K=2$, and $K=1$, respectively, are employed. To be able to compare the effect of the immersion-based coding algorithm, the immersed optimizers based on the immersion coding given in Proposition \ref{proposition1}, target SGD, target Momentum, and target Adam optimizers are employed for training SIFL models.
\end{itemize}
We set the number of clients to $N_c=10$. Our implementation uses Keras with a Tensorflow backend on an HP laptop with A100 GPU and 16 GB RAM.\\
\indent We implement various FL algorithms through standard FL (FL), the SIFL Method (SIFL M1) given in Algorithm \ref{alg:one}, and the extended SIFL Method (SIFL M2) in Algorithm \ref{alg:two}. To be able to implement SIFL M1 and SIFL M2, the variables of the encoding mechanisms and the target optimizer are designed by selecting small full-rank matrices $\Pi_1$ and $\Pi_2$ with appropriate dimensions. We compute the base $N_1$ of the kernel of $\Pi_{1}^{L}$ and the base $N_2$ of the kernel of $\Pi_{2}^{R}$. The random processes $R_1^t$ and $R_2^t$ are defined at every round as multivariate Gaussian variables with large covariances. The immersed dimensions of model parameters are shown in Table \ref{tab:params}.\\
To calculate DP guarantees for local and global models according to Theorem \ref{theoremLaplace}, first we determine the sensitivities of local and global models $\Delta_1^{w_i}$ and $\Delta_1^{w}$, which according to Remark \ref{remark3} and considering the clipping threshold $C=1000$, number of clients $N_c=10$, and the size of dataset $|\mathcal{D}|=60000$ and local datasets $|\mathcal{D}|_i=6000$, can be calculated as $\Delta_1^{w_i}=\frac{2C}{|\mathcal{D}_i|}=0.33$ and $\Delta_1^{w}=\frac{2C}{|\mathcal{D}|}=0.033$. Then, based on Theorem \ref{theoremLaplace}, considering encoding matrices $\Pi_1$ and $\Pi_2$, with $||\Pi_1^j||_1=10^{-3}$, $||\Pi_2^R||_2=10^3$, $||N_1^j||_2=10^3$, $||\Pi_2^m||_2=10^{-3}$and Laplace processes $R_1^t \sim \operatorname{Laplace }(\mathbf{0}, \sigma_1 I)$ and $R_2^t \sim \operatorname{Laplace }(\mathbf{0}, \sigma_2 I)$ with standard deviations $\sigma_1=\sigma_2=10^3$, the $\Tilde{\epsilon}$ and $\epsilon^{\prime}$-DP guarantees with $\Tilde{\epsilon}=1e-12$ and $\epsilon^{\prime}=1e-13$ are warranted for each elements of local and global models, which are very high-levels of DP-guarantee. Note that since in this method, the privacy noises are removed by the server and clients, they do not need to be small. Therefore, high levels of DP guarantees are achieved without distorting the performance of the FL model.
\begin{table}[t]
  \caption{Dimensions of model parameters.}
  \centering
  \begin{tabular}{llll}
    \toprule
    \textbf{Model} & $n$ & $\Tilde{n}$&  $n^\prime$\\
    \midrule
    MLP & 199,210& 199,411 & 398,822\\
    CNN & 1,199,882 & 1,200,011& 2,400,022\\
    CNN2 & 582,026& 582,539 & 1,165,078\\
    \bottomrule
  \end{tabular}
  \label{tab:params}
\end{table}
\subsection{Performance Evaluation}
\begin{figure}[hb]
\centering
\subfloat{\includegraphics[width = 1.6in]{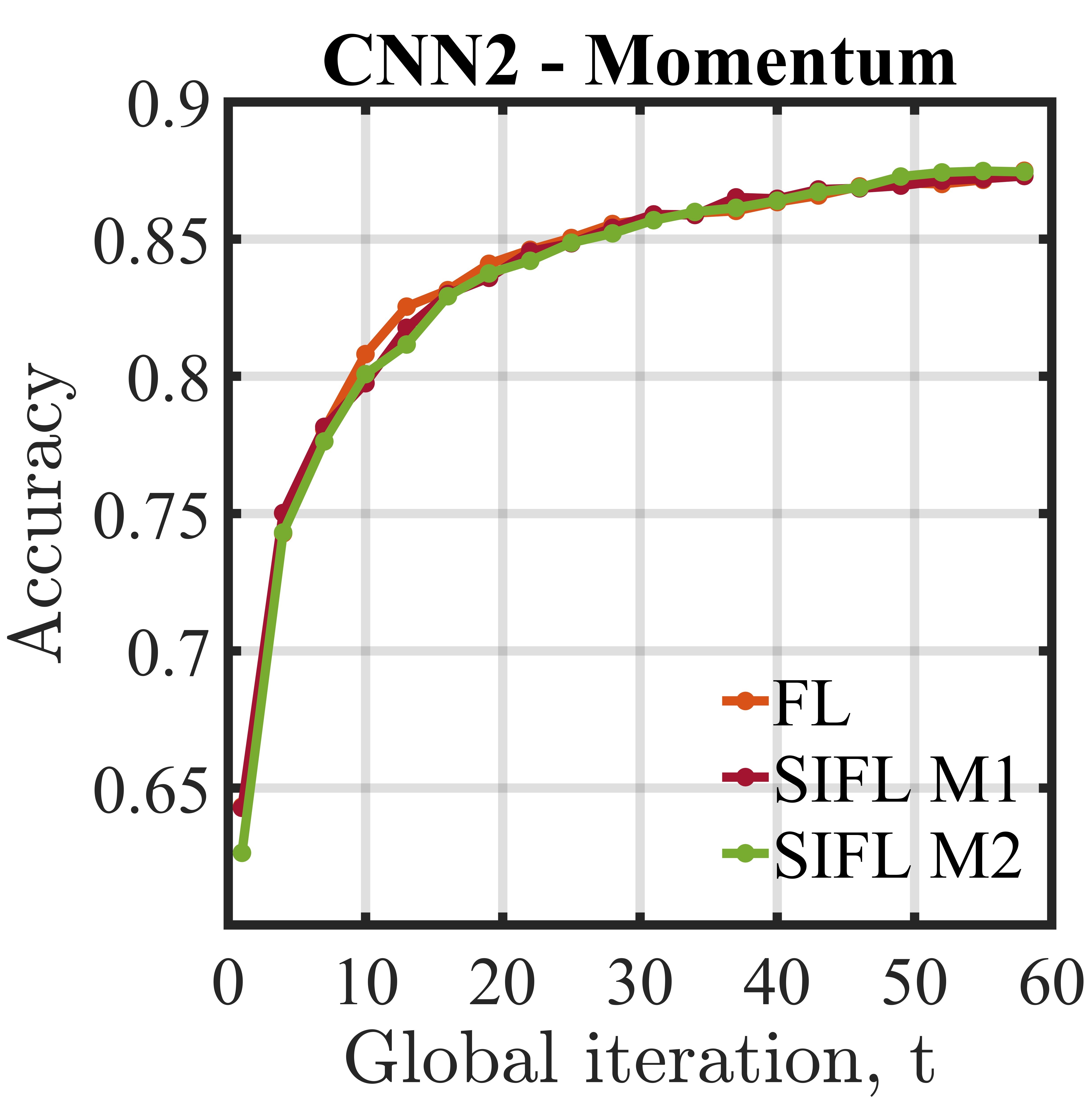}} 
\subfloat{\includegraphics[width = 1.6in]{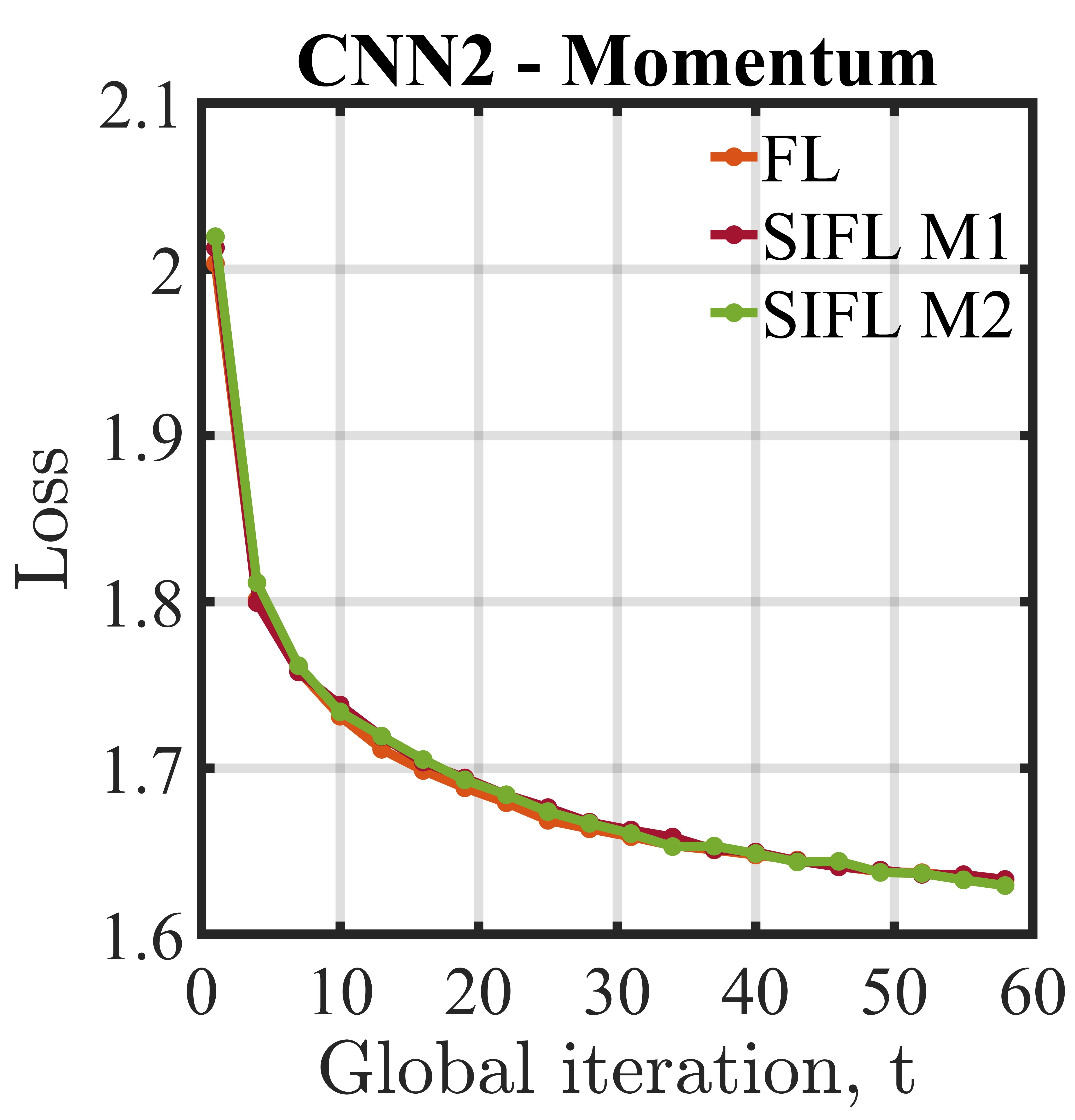}}  
\caption{The comparison of the accuracy and loss of FL with and without privacy for the Fashion-MNIST database.}
\label{fig:FMNIST}
\end{figure}

\begin{figure*}[ht]
\centering
\subfloat[MLP-ADAM]{\includegraphics[width = 1.9in]{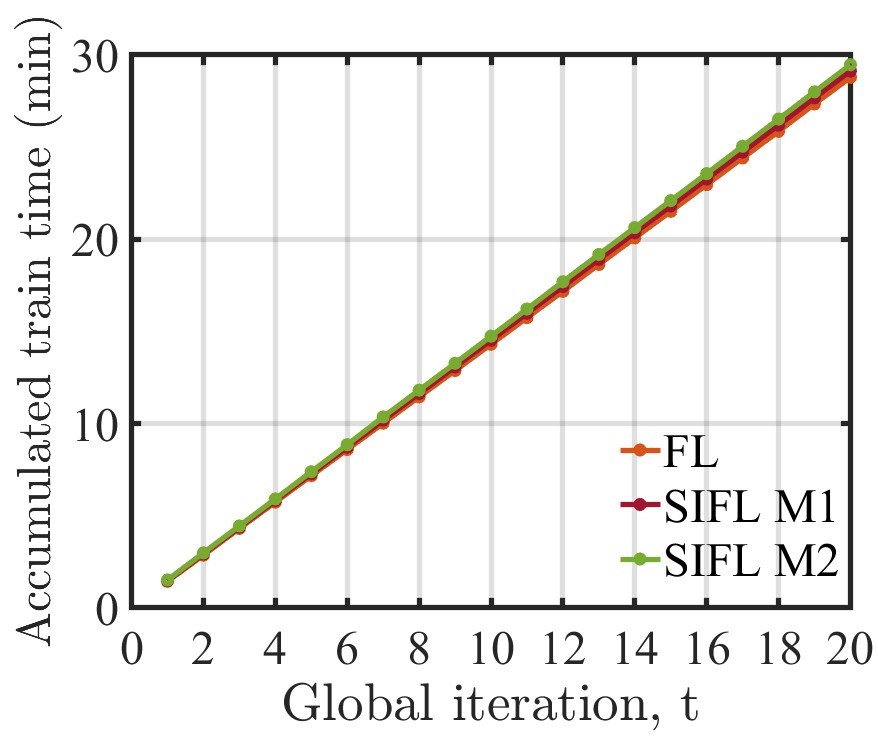}} 
\subfloat[CNN2-Momentum]{\includegraphics[width = 1.9in]{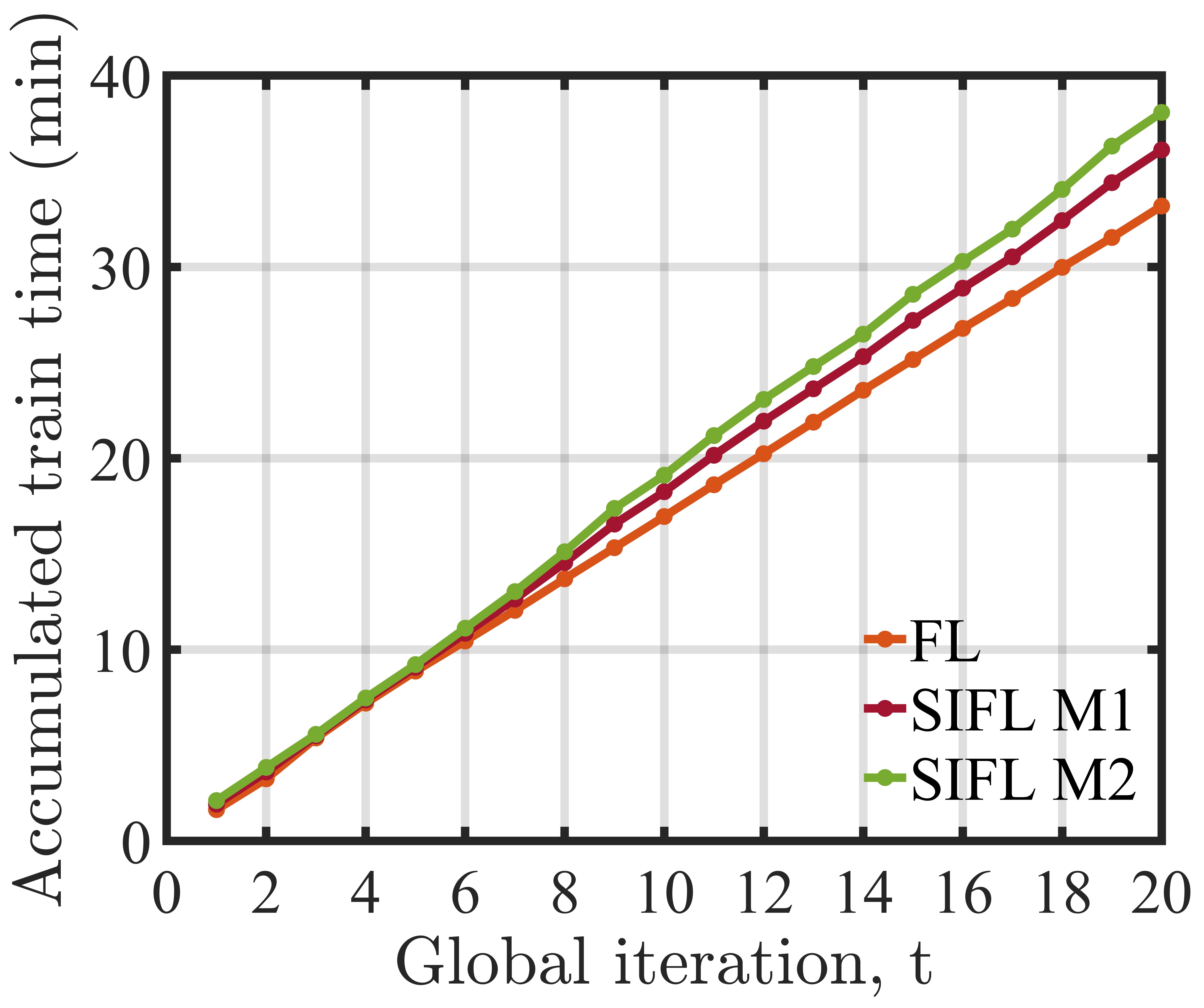}}
\subfloat[CNN-SGD]{\includegraphics[width = 1.9in]{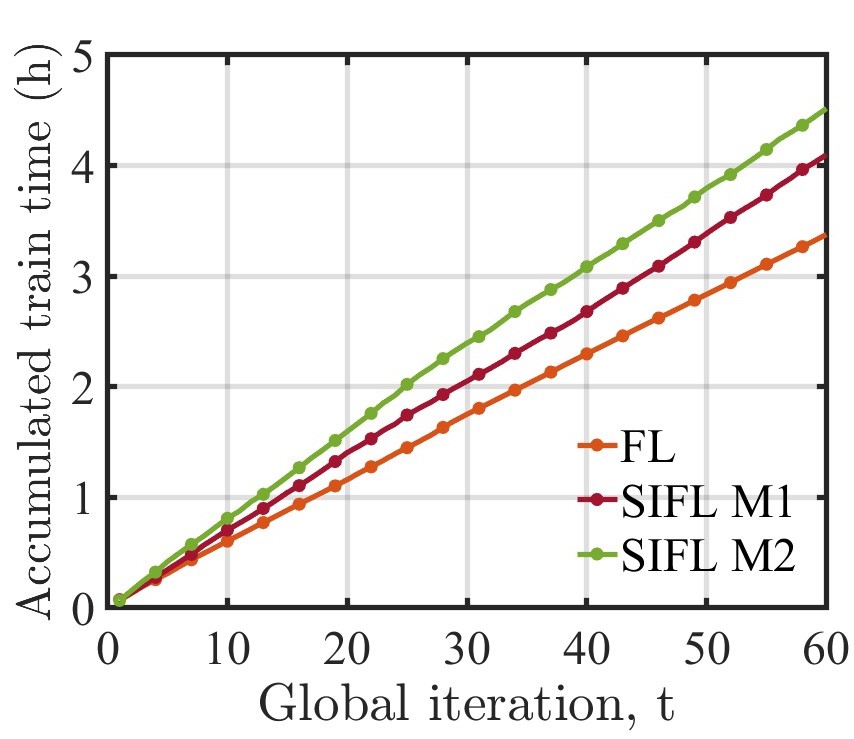}}  
\caption{The comparison of the training time of FL with and without the proposed privacy mechanism.}
\label{fig:timecost}
\end{figure*}

\vspace{-2mm}
\indent The comparison of training accuracy of the standard FL algorithm and the proposed SIFL M1 and SIFL M2 are shown in Figure \ref{fig:accuracy}. The test accuracies are shown for different models (MLP and CNN), using different optimizers (SGD, Adam, and momentum) and their equivalent target optimizers for SIFL M1 and SIFL M2. The comparison of the accuracy of the FL algorithm for the Fashion-MNIST database and CNN2 model is shown in Figure \ref{fig:FMNIST}. As can be seen, the SIFL M1 and SIFL M2 accuracy is almost the same as the accuracy with no privacy setting in all scenarios, which shows that SIFL can integrate a cryptographic method in the FL system without sacrificing model accuracy and convergence rate. Therefore, there is no need to make a trade-off between privacy and FL performance.\\
\indent In Figures \ref{fig:timecost}, we investigate the effect of the encoding and decoding operations in SIFL M1 and SIFL M2 on FL training time. As can be seen, for the MLP model, where the number of model parameters is smaller, the additional training time compared to the training time of the original FL is negligible. However, by increasing the number of model parameters in CNN models, the training time increases. The reason is that by increasing the number of model parameters, the size of multiplicative matrices in privacy mechanisms $\Pi_1$ and $\Pi_2$ increases. In addition, the model parameters are flattened at every iteration to become a vector to be able to apply privacy mechanisms, which takes more time.\\
\begin{figure}[htb]\centering
\includegraphics[width=.94\linewidth]{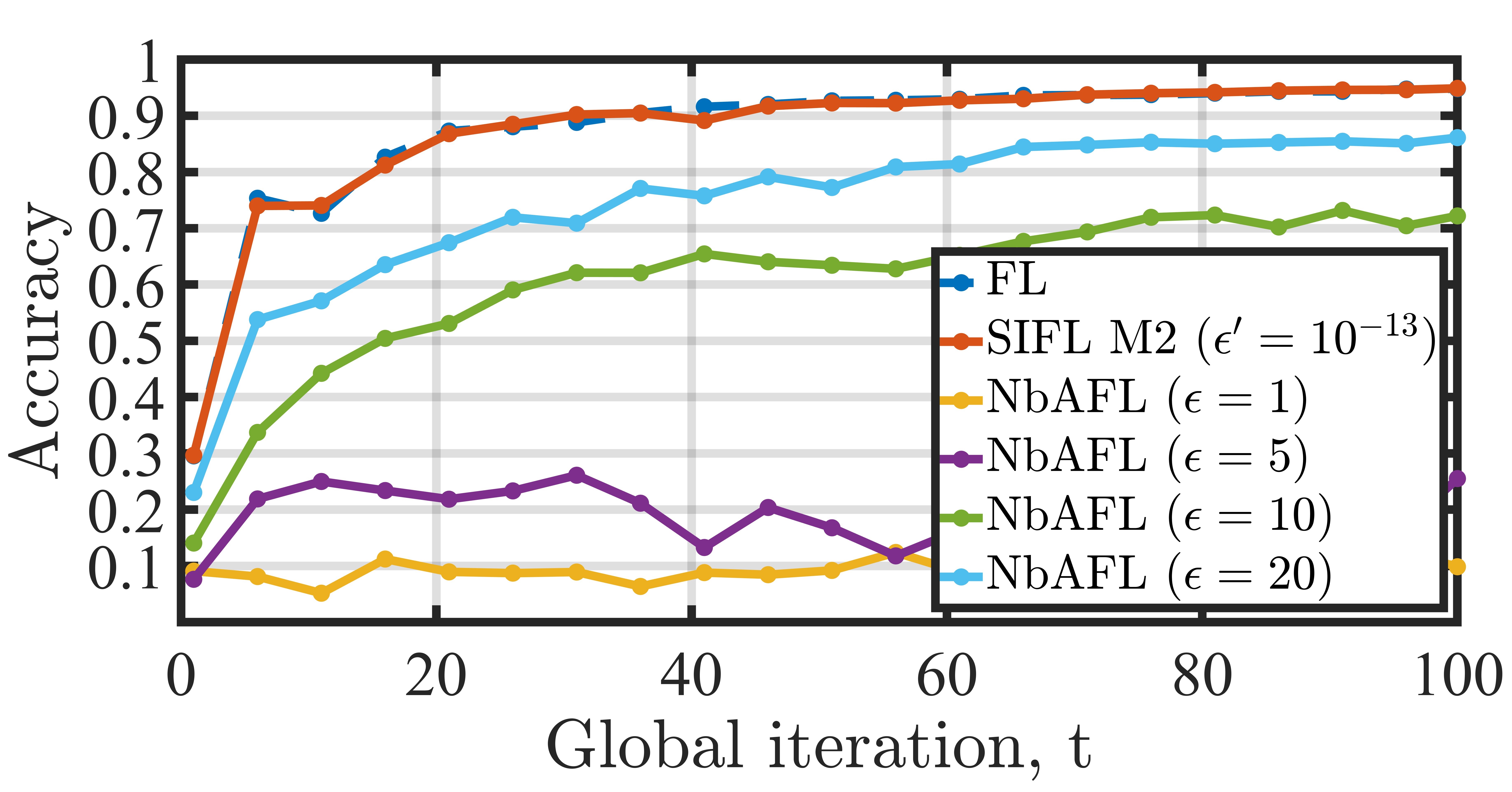}
  \caption{The comparison of the accuracy of FL, SIFL M2, and NbAFL for various privacy levels $\epsilon=1,5,10,20$.}\label{dpcompare}
\end{figure}
\indent We compare the accuracy of our proposed scheme with a differential privacy-based FL algorithm in Figure \ref{dpcompare}, in which the distortion induced by the DP noises is not removed from the model. The federated learning algorithm with differential privacy proposed in \cite{wei2020federated}, namely, noising before model aggregation FL (NbAFL) is employed to compare the accuracy. It should be noted that to be able to calculate the sensitivity of model parameters in federated learning with differential privacy algorithms, a clipping technique is employed to ensure that $||w^t_i|| \le C$ with clipping threshold $C$. In standard FL with DP algorithms, if the clipping threshold $C$ is too small, clipping destroys the intended gradient direction of parameters, and if it is too large, it forces to add more noise to the parameters because of its effect on the sensitivity. However, in SIFL, since the server and the clients can remove the distortion induced by the privacy noises, they do not need to be small. Therefore, we can choose a large clipping threshold to avoid distorting the FL performance. Hence, in this implementation, the clipping threshold for NbAFL is $C=10$, and for SIFL is $C=1000$. Other parameters in this implementation are $N_{\mathcal{C}}=50$ and batch size is $1024$, CNN2 model with SGD optimizer and learning rate $0.1$ are employed. We measure the model accuracy of NbAFL in a given DP levels $(\epsilon, \delta)$ for $\delta=1e-5$ and $\epsilon=\{1,5,10,20\}$.
To implement SIFL algorithm, we used Gaussian additive noises $R_1^t\sim \mathcal{N}(\mathbf{0}, \sigma_1 I)$ and $R_2^t \sim \mathcal{N}(\mathbf{0}, \sigma_2 I)$ with standard deviations $\sigma_1=\sigma_2=10^3$. According to Theorem \ref{theoremGaussian} for calculating the DP guarantees for Gaussian noises, considering $||\Pi_1^j||_2=10^{-3}$, $||\Pi_2^R||_2=10^3$, $||N_1^j||_2=10^3$, $\sigma_1=\sigma_2=10^3$, sensitivities of local and global models $\Delta_1^{w_i}=0.33$ and $\Delta_1^{w}=0.033$, the $(\Tilde{\epsilon},\Tilde{\delta})$ and $({\epsilon}^{\prime},{\delta}^{\prime})$-DP guarantees for each element of local and global models can be achieved with $\Tilde{\epsilon}=1e-11$, $\Tilde{\delta}=1e-5$, ${\epsilon}^{\prime}=1e-13$, and ${\delta}^{\prime}=1e-5$ which are very high levels of DP guarantee. As can be seen in Figure \ref{dpcompare}, a higher level of DP guarantee in a standard DP-based FL algorithm like NbAFL would affect the accuracy and convergence rate significantly due to the distorting noises, while the proposed SIFL algorithm can provide a very high level of DP guarantee without losing the model accuracy.
\vspace{-1mm}
\section{Conclusions}\label{sec7}
\vspace{-2mm}
In this paper, we have developed a System Immersion Federated Learning, SIFL, as a privacy-preserving FL framework built on the synergy of random coding and system immersion tools from control theory to protect privacy of the clients' data in federated learning. The core idea is to treat the Gradient descent optimization algorithm employed in the standard FL process as a dynamical system that we seek to immerse into a higher-dimensional system. We have devised a synthesis procedure to design the dynamics of a coding scheme for privacy and an immersed higher-dimensional optimization algorithm called target optimizer such that model parameters of the standard optimization algorithm are immersed/embedded in its parameters, and it operates on randomly encoded higher-dimensional model parameters. Random coding was formulated at the server as a random change of coordinates that maps the original private parameters of the FL model to a higher-dimensional space. Such coding enforces that the target optimization algorithm converges to an encoded higher-dimensional version of the model parameters of the original optimization algorithm that can be decoded at the server after model aggregation.\\
\indent SIFL provides the same accuracy and convergence rate as the standard FL  (i.e., when no coding is employed to protect against data inference), reveals no information about clients' data, can be applied to large-scale models, is computationally efficient, and offers any desired level of differential privacy without degrading the FL accuracy and performance. The simulation results of SIFL are presented to illustrate the performance of our tool. These results demonstrate that SIFL provides the same accuracy and convergence rate as the standard FL with a negligible computation cost.
\bibliographystyle{elsarticle-num} 
\bibliography{conference_101719}
\vspace{-1mm}
\section{Appendix}
\vspace{-2mm}
\subsection{Proof of Theorem \ref{theoremGaussian}}\label{Prooftheoremgaussian}
\vspace{-2mm}
We start with the privacy guarantee for ${w}^t_i$. According to Definition \ref{defSensitivity}, given the adjacent local databases $\mathcal{D}_i, \mathcal{D}^{\prime}_i \in \mathcal{X}_i$, the sensitivity of ${w}^t_i$ can be defined as follows:
\begin{equation}\label{sensitivityy2}
    \Delta^{w_i}_2=\sup _{\mathcal{D}_i, \mathcal{D}^{\prime}_i}\left\| {w}^t_i \right(\mathcal{D}_i \left) - {w}^t_i\left(\mathcal{D}^{\prime}_i \right)\right\|_2.
\end{equation}
For simplicity, we show ${w}^t_i(\mathcal{D}_i)$ and ${w}^t_i(\mathcal{D}^{\prime}_i)$ by ${w}^t_i$ and ${{w}^t_i}^{\prime}$. Because $R_1^t \sim \mathcal{N}(\mathbf{0}, \sigma_1 I)$, and given the privacy encoding mechanisms $\tilde{w}^t_i= \Pi_1 {w}^t_i + b_1^t$, with $b_1^t=N_1 R_1^t \Pi_2^R$ in the extended SIFL method, each element of $\tilde{w}^t_i$ follows a Gaussian distribution as $\tilde{w}^t_{i,j} \sim \mathcal{N}(\Pi_{1}^{j} {w}^t_{i},||N_1^{j}||_2 \sigma_1 ||\Pi_2^R||_2)$, where $\Pi_1^{j}$ and $N_1^{j}$ are the $j$-th rows of $\Pi_1$ and $N_1$, respectively. We have
\begin{equation}
\begin{aligned}
& \mathbb{P}\left(\tilde{w}^t_{i,j}(\mathcal{D}_i)\right)=\frac{1}{\left(2 \pi \bar{\sigma}^2\right)^\frac{1}{2}} \int_{{\mathcal{S}}} e^{-\frac{||p-\Pi_{1}^{j} {w}^t_{i}||_2^2}{2 (||N_1^{j}||_2 \sigma_1 ||\Pi_2^R||_2)^2}} d p\\
&= \frac{1}{\left(2 \pi \bar{\sigma}^2\right)^\frac{1}{2}} \int_{\mathcal{S}} e^\frac{-||p-\Pi_{1}^{j} {{w}^t_i}^{\prime}-\Pi_{1}^{j} v||_2^2}{\left(2 (||N_1^{j}||_2 \sigma_1 ||\Pi_2^R||_2)^2\right)} d p \\
&= \frac{1}{\left(2 \pi \bar{\sigma}^2\right)^\frac{1}{2}} \int_{\mathcal{S}} e^{-\frac{||p-\Pi_{1}^{j} {{w}^t_i}^{\prime}||_2^2}{2 (||N_1^{j}||_2 \sigma_1 ||\Pi_2^R||_2)^2}} 
  e^{\frac{2(p-\Pi_{1}^{j} {{w}^t_i}^{\prime}) \Pi_{1}^{j} v-||\Pi_{1}^{j} v||_2^2}{2 (||N_1^{j}||_2 \sigma_1 ||\Pi_2^R||_2)^2}} d p \\
&= \frac{1}{\left(2 \pi \bar{\sigma}^2\right)^\frac{1}{2}} \int_{{\mathcal{S}} \cap A_{\epsilon}} e^{-\frac{||p-\Pi_{1}^{j} {{w}^t_i}^{\prime}||_2^2}{2 (||N_1^{j}||_2 \sigma_1 ||\Pi_2^R||_2)^2}} 
 e^{\frac{2(p-\Pi_{1}^{j} {{w}^t_i}^{\prime}) \Pi_{1}^{j} v-||\Pi_{1}^{j} v||_2^2}{2 (||N_1^{j}||_2 \sigma_1 ||\Pi_2^R||_2)^2}} d p \\
&+\frac{1}{\left(2 \pi \bar{\sigma}^2\right)^\frac{1}{2}}\int_{{\mathcal{S}} \cap A_{\epsilon}^c} e^{-\frac{||p-\Pi_{1}^{j} {w}^t_{i}||_2^2}{2 (||N_1^{j}||_2 \sigma_1 ||\Pi_2^R||_2)^2}} d p,
\end{aligned}
\end{equation}
where $v:={w}^t_i -{{w}^t_i}^{\prime}$, $\bar{\sigma}=(||N_1^{j}||_2 \sigma_1 ||\Pi_2^R||_2)$, \\$A_{\epsilon}=\left\{p \in \mathbb{R}: \frac{2(p-{{w}^t_i}^{\prime}) \Pi_{1}^{j} v-||\Pi_{1}^{j} v||_2^2}{2 (||N_1^{j}||_2 \sigma_1 ||\Pi_2^R||_2)^2} \leq \tilde{\epsilon}\right\}$ and $A_{\epsilon}^c$ denotes its complement. By the definition of $A_{\epsilon}$, the first term of the last expression is bounded by
\begin{equation}
\frac{e^{\tilde{\epsilon}}}{\left(2 \pi \bar{\sigma}^2\right)^\frac{1}{2}}\int_{\mathcal{S}} e^{-\frac{||p-{{w}^t_i}^{\prime}||_2^2}{2 (||N_1^{j}||_2 \sigma_1 ||\Pi_2^R||_2)^2}} d p=e^{\tilde{\epsilon}} \mathbb{P}\left(\tilde{w}^t_{i,j}(\mathcal{D}^{\prime}_i) \in {\mathcal{S}} \right).
\end{equation}
The second integral term is bounded by
\begin{equation}\scalebox{0.97}{$
\frac{1}{\left(2 \pi \bar{\sigma}^2\right)^\frac{1}{2}} \int_{\mathbb{R}} e^{-\frac{||p-\Pi_{1}^{j} {w}^t_{i}||_2^2}{2 (||N_1^{j}||_2 \sigma_1 ||\Pi_2^R||_2)^2}} 1_{\{2(p-\Pi_{1}^{j}{{w}^t_i}^{\prime}) \Pi_{1}^{j} v \geq||\Pi_{1}^{j} v||_2^2+2 {\tilde{\epsilon}} {\bar{\sigma}}^2\}} d p,$}
\end{equation}
which, after the change of variable $y=(p-\Pi_{1}^{j} {w}^t_{i}) / \bar{\sigma}$, can be rewritten
\begin{equation}
\begin{aligned}
& \frac{1}{(2 \pi)^\frac{1}{2}} \int_{\mathbb{R}} e^{-\frac{||y||_2^2}{2}} 1_{\left\{2(\bar{\sigma} y+\Pi_{1}^{j} v) \Pi_{1}^{j} v \geq ||\Pi_{1}^{j} v||_2^2+2 {\tilde{\epsilon}} {\bar{\sigma}}^2\right\}} d y \\
= & \frac{1}{(2 \pi)^\frac{1}{2}} \int_{\mathbb{R}} e^{-\frac{||y||_2^2}{2}} 1_{\left\{y \geq \frac{{\tilde{\epsilon}}\bar{\sigma}}{||\Pi_{1}^{j} v||_2}-\frac{||\Pi_{1}^{j} v||_2}{2 \bar{\sigma}}\right\}} d y.
\end{aligned}
\end{equation}
This last expression is $\mathbb{P}\left(Y \geq \frac{{\tilde{\epsilon}}\bar{\sigma}}{||\Pi_{1}^{j} v||_2}-\frac{||\Pi_{1}^{j} v||_2}{2 \bar{\sigma}}\right) \leq \tilde{\delta}$, for $Y \sim \mathcal{N}\left(0, 1\right)$. 
We are then led to set $\bar{\sigma}$ sufficiently large so that $\mathbb{P}\left(Y \geq {\tilde{\epsilon}} \bar{\sigma} /||\Pi_{1}^{j} v||_2-||\Pi_{1}^{j} v||_2 / 2\bar{\sigma}\right) \leq \tilde{\delta}$, that is, $Q\left({\tilde{\epsilon}} \bar{\sigma} /||\Pi_{1}^{j} v||_2-||\Pi_{1}^{j} v||_2 / 2 \bar{\sigma}\right) \leq \tilde{\delta}$. Because $Q^{-1}$ is monotonically decreasing, we have the condition $\frac{{\tilde{\epsilon}} \bar{\sigma}}{||\Pi_{1}^{j} v||_2}-\frac{||\Pi_{1}^{j} v||_2}{2 \bar{\sigma}} \geq Q^{-1}(\tilde{\delta})$, which is equivalent to
\begin{equation}\label{ineq gauss}
{\bar{\sigma}}^2- {\bar{\sigma}} \frac{||\Pi_{1}^{j} v||_2}{{\tilde{\epsilon}}} Q^{-1}(\tilde{\delta})-\frac{||\Pi_{1}^{j} v||_2^2}{2 {\tilde{\epsilon}}} \geq 0.
\end{equation}
From Definition \ref{defSensitivity}, \eqref{ineq gauss} can be converted to:
\begin{equation}\label{ineq gauss2}
{{\bar{\sigma}}}^2- {\bar{\sigma}} \frac{||\Pi_{1}^{j}||_2\Delta^w_2
}{{\tilde{\epsilon}}} Q^{-1}(\tilde{\delta})-\frac{||\Pi_{1}^{j}||_2^2(\Delta^w_2)^2
}{2 {\tilde{\epsilon}}} \geq 0.
\end{equation}
Substituting $\bar{\sigma}=(||N_1^{j}||_2 \sigma_1 ||\Pi_2^R||_2)$ into \eqref{ineq gauss2}, to have $(\Tilde{\epsilon},\Tilde{\delta})$-DP guarantee for local models, $\Pi_1$, $\Pi_2$, $N_1$, and $\sigma_1$ should be designed to satisfy the following inequality:
\begin{equation}\label{ineq gauss22}
\begin{aligned}
&(||N_1^{j}||_2 \sigma_1 ||\Pi_2^R||_2)^2- \frac{|\Pi_{1}^{j}|_2\Delta^w_2
}{\tilde{\epsilon}} Q^{-1}(\tilde{\delta})(||N_1^{j}||_2 \sigma_1 ||\Pi_2^R||_2)\\
&\quad\quad-\frac{||\Pi_{1}^{j}||_2^2(\Delta^w_2)^2
}{2 \tilde{\epsilon}} \geq 0.
\end{aligned}
\end{equation}
Following the same reasoning, we define the sensitivity of the global model as $\Delta^{w}_2=\sup _{\mathcal{D}_i, \mathcal{D}^{\prime}_i}\left\| {w}^t \right(\mathcal{D}_i \left) - {w}^t\left(\mathcal{D}^{\prime}_i \right)\right\|_2$. For the extended SIFL method, it can be shown that each element of the encoded global model ${w^{\prime}}^t = \Pi_1 {w}^t \Pi_2 + B_1^t + \Pi_1\Pi_1^L B^t_2$, ${w^{\prime}}^t _{j,m}$, with $B_1^t= N_1 R_1^t$ and $B_2^t=R_2^t N_2$ is $(\epsilon^{\prime}, \delta^{\prime})$-Differentially private in \eqref{dpcondition}) for:
\begin{equation}\label{ineq gauss global}
\begin{aligned}
&(||N_1^{j}||_2 \sigma_1 +||N_2^{m}||_2 \sigma_2)^2 -\frac{||\Pi_{1}^{j}||_2^2(\Delta^w_2)^2 ||\Pi_{2}^{m}||^2_2
}{2 {\epsilon}^{\prime}}\\
&\quad \quad \quad - \frac{||\Pi_{1}^{j}||_2\Delta^w_2
||\Pi_{2}^{m}||_2}{{\epsilon}^{\prime}} Q^{-1}({\delta}^{\prime})(||N_1^{j}||_2 \sigma_1 \\&\quad \quad \quad +||\Pi^j_1||_2 ||\Pi^L_1||_2 ||N_2||_2 \sigma_2) \geq 0,
\end{aligned}
\end{equation}
for all $j \in \{1,2,\dots,\tilde{n}\}$ and $m \in \{1,2,\dots,p\}$, where $N_2^{m}$ and $\Pi_{2}^{m}$ are the $m^{th}$ column of $N_2$ and $m^{th}$ element of vector $\Pi_{2}$, respectively.\hfill $\blacksquare$
\end{document}